\documentclass[twocolumn]{aastex63}
\usepackage{amsmath,amssymb,amstext}
\usepackage{gensymb}
\usepackage{multirow}
\usepackage{rotating}
\usepackage{tabularx}

\graphicspath{{./}{figures/}}

\newcommand{\spitzer}{{\textit{Spitzer}}}
\newcommand{\hst}{{\textit{HST}}}
\newcommand{\kepler}{{\textit{Kepler}}}
\newcommand{\tess}{{\textit{TESS}}}

\defcitealias{MayKomacek2021}{May \& Komacek et al.}

\shorttitle{Spitzer Phase Curves Galore}
\shortauthors{May et al.}

\begin{document}

\title{A New Analysis of 8 \spitzer\ Phase Curves and Hot Jupiter Population Trends: \\ Qatar-1b, Qatar-2b, WASP-52b, WASP-34b, and WASP-140b}

\correspondingauthor{E. M. May}
\email{Erin.May@jhuapl.edu}

\author[0000-0002-2739-1465]{E. M. May}

\author[0000-0002-7352-7941]{K. B. Stevenson}
\affiliation{Johns Hopkins APL, 11100 Johns Hopkins Rd, Laurel, MD 20723, USA}

\author[0000-0003-4733-6532]{Jacob L. Bean}
\affiliation{Department of Astronomy \& Astrophysics, University of Chicago, 5640 S. Ellis Avenue, Chicago, IL 60637, USA}

\author[0000-0003-4177-2149]{Taylor J. Bell}
\affiliation{Department of Physics, McGill University, Montreal, QC H3A 2T8, Canada}
\affiliation{BAER Institute, NASA Ames Research Center, Moffet Field, CA 94035, USA}

\author[0000-0001-6129-5699]{Nicolas B. Cowan}
\affiliation{Department of Physics, McGill University, Montreal, QC H3A 2T8, Canada}
\affiliation{Department of Earth and Planetary Sciences, McGill University, Montreal, QC H3A 0E8, Canada}

\author[0000-0003-4987-6591]{Lisa Dang}
\affiliation{Department of Physics, McGill University, Montreal, QC H3A 2T8, Canada}

\author[0000-0002-0875-8401]{Jean-Michel Desert}
\affiliation{Anton Pannekoek Institute for Astronomy, University of Amsterdam, 1090 GE Amsterdam, Netherlands}

\author[0000-0002-9843-4354]{Jonathan J. Fortney} 
\affiliation{Department of Astronomy \& Astrophysics, University of California, Santa Cruz, CA 95064, USA}

\author[0000-0001-9887-4117]{Dylan Keating} 
\affiliation{Department of Physics, McGill University, Montreal, QC H3A 2T8, Canada}

\author[0000-0002-1337-9051]{Eliza M.-R. Kempton}
\affiliation{Department of Astronomy, University of Maryland, College Park, MD 20742, USA}

\author[0000-0002-9258-5311]{Thaddeus D. Komacek}
\affiliation{Department of Astronomy, University of Maryland, College Park, MD 20742, USA}

\author[0000-0002-8507-1304]{Nikole K. Lewis}
\affiliation{Department of Astronomy and Carl Sagan Institute, Cornell University, 122 Sciences Drive, Ithaca, NY 14853, USA}

\author[0000-0003-4241-7413]{Megan Mansfield}
\altaffiliation{NHFP Sagan Postdoctoral Fellow}
\affiliation{Steward Observatory, University of Arizona, Tucson, AZ 85719, USA}

\author[0000-0002-4404-0456]{Caroline Morley}
\affiliation{Department of Astronomy, University of Texas at Austin, Austin, TX 78712, USA}

\author[0000-0001-9521-6258]{Vivien Parmentier}
\affiliation{Department of Physics (Atmospheric, Oceanic and Planetary Physics), University of Oxford, Parks Rd, Oxford, OX1 3PU, UK}

\author[0000-0003-3963-9672]{Emily Rauscher}
\affiliation{Department of Astronomy, University of Michigan, Ann Arbor, MI 48109, USA}

\author[0000-0002-0919-4468]{Mark R. Swain}  
\affiliation{Jet Propulsion Laboratory (JPL), California Institute of Technology, Pasadena, CA 91109, USA}

\author[0000-0001-7547-0398]{Robert T. Zellem}
\affiliation{Jet Propulsion Laboratory (JPL), California Institute of Technology, Pasadena, CA 91109, USA}

\author{Adam Showman}
\altaffiliation{Deceased.}
\affiliation{Lunar and Planetary Laboratory, University of Arizona, Tucson, AZ, 85721, USA}

\begin{abstract}
With over 30 phase curves observed during the warm \spitzer\ mission, the complete data set provides a wealth of information relating to trends and three-dimensional properties of hot Jupiter atmospheres. In this work we present a comparative study of seven new \spitzer\ phase curves for four planets with equilibrium temperatures of T$_{eq}\sim$ 1300K: Qatar-2b, WASP-52b, WASP-34b, and WASP-140b, as well as the reanalysis of the 4.5 $\micron$ Qatar-1b phase curve due to the similar equilibrium temperature. In total, five 4.5 $\micron$ phase curves and three 3.6 $\micron$ phase curves are analyzed here with a uniform approach. Using these new results, in combination with literature values for the entire population of published \spitzer\ phase curves of hot Jupiters, we present evidence for a linear trend of increasing hot spot offset with increasing orbital period, as well as observational evidence for two classes of planets in apparent redistribution vs. equilibrium temperature parameter space, and tentative evidence for a dependence of hot spot offset on planetary surface gravity in our $\sim$ 1300 K sample. We do not find trends in apparent heat redistribution with orbital period or gravity. Non-uniformity in literature \spitzer\ data analysis techniques precludes a definitive determination of the sources or lack of trends.
\end{abstract}

\section{Introduction} \label{intro}
Phase curve observations are key to studying the strength and type of circulation in the atmosphere, as well as being the primary way to probe the nightside of a planet. In addition to a direct measure of the heat transport efficiency of the atmosphere - infrared phase curve observations probe transmission and emission over the course of the planet's orbit. In total, a phase curve provides the most comprehensive view of a given exoplanet's global atmospheric state. The scientific potential of phase curves do not stop with single planet studies: observed population trends can tell us how atmospheric dynamics vary with key planetary parameters, constraining and differentiating between various atmospheric models. 

The days-long orbital periods of hot Jupiters require space-based observations of their phase curves. Although multiple space telescopes exist to perform these observations, \spitzer's access to infrared wavelengths (at which planets emit thermal radiation) complements the wavelengths of optical to near infrared observatories like \hst, \kepler, and \tess\ which are more sensitive to reflected light \footnote{While \hst's WFC3 does probe emission for many hot planets, it is at different atmospheric regions than that of \spitzer}. This unique access to the infrared made \spitzer\ a popular choice for exoplanet observations - over the lifetime of the warm \spitzer\ mission, phase curves of over 30 planets were observed, totalling over 5 dozen data sets between the 3.6 and 4.5 $\micron$ channels. With notable recent exceptions (see below discussion), many analyses of these phase curve observations have been single planet studies with individualized reduction techniques. 

\spitzer\ InfraRed Array Camera \citep[IRAC,][]{Fazio2004} observations at 3.6 and 4.5 $\micron$ are primarily affected by the intrapixel sensitivity effect, where the flux measured varies on the order of a few percent as a centroid drifts within a single pixel, or sometimes to neighboring pixels \citep[e.g.][]{Reach2005,Charbonneau2005,Knutson2008,Knutson2009, Ingalls2012}. Previous \spitzer\ analyses of exoplanet phase curves have used various methods to remove the intrapixel effect. It is these single-planet studies with non-uniform methods that limit the comparative exoplanetology possible from the entire \spitzer\ phase curve population. Analysis of HD 189733b and HD 209458b used a Gaussian regression of the centroid location and a noise parameter \citep{Knutson2012,Zellem2014}; Pixel-level decorrelation (PLD) was used for HAT-P-7b, HD 149026b, WASP-19b, WASP-14b and WASP-33b \citep{Wong2015, Wong2016, Zhang2018}; Detrending via Legendre Polynomials for WASP-18b \citep{Maxted2013}; while a form of a intrapixel sensitivity mapping method (e.g. Ballard Map, \citealt{Ballard2010}; BLISS map, \citealt{Stevenson2012}; or a unique method) was used for HAT-P-7b, WASP-19b, WASP-14b, WASP-43b, WASP-103b, WASP-76b, Qatar-1b, and KELT-9b (\citealt{Cowan2012, Wong2015, Wong2016, Stevenson2017, Mendonca2018, Kreidberg2018, Bell2019, Keating2020, Mansfield2020, May2020}; \citetalias{MayKomacek2021}, \citeyear{MayKomacek2021}). \cite{Krick2016} used the calibration star BD+67 1044 to calibrate a sparsely sampled phase curve of WASP-14b, a similar method to our fixed sensitivity map we discuss below. \cite{Bell2021} presented the first uniform re-analysis of the above planets, in addition to MASCARA-1b and KELT-16b, comparing four different systematic models for all planets (BLISS mapping, PLD, polynomials, and a Gaussian process). 

\cite{Bell2021} found that the BLISS (Bilinearly Interpolated Subpixel Sensitivity) mapping method performs best for most phase curves and in \cite{May2020} we presented an update to that method at 4.5 $\micron$ to enable a more uniform approach for systematic detrending. This update applies a fixed intrapixel sensitivity correction map generated with \spitzer\ IRAC calibration data rather than the standard method of self-calibration. This minimizes differing residual systematics between data sets to enable comparative studies. Naturally, when moving to population level studies, it is of key importance that we determine if our measured trends are astrophysical in nature or if there are unconstrained systematics in our data sets.   

Numerous trends have been predicted and inferred from phase curve observations of hot Jupiters, primarily as a function of equilibrium and/or irradiation temperature\footnote{The irradiation temperature is related to the equilibrium temperature as $T_{\rm{irr}} = \sqrt{2}T_{\rm{eq}}$.}. \cite{PerezBecker2013} and \cite{Komacek2016} explore the observational trend of increasing day-night contrast with increasing equilibrium temperature with 3D models, suggesting that these trends arise due to the decreasing ratio of radiative cooling timescales to day-night wave propagation timescales. \cite{Komacek2017} take these predictions at varying atmospheric drag timescales and directly compare them to available observational data, finding models best match data at higher equilibrium temperatures - notably where more recent work suggests clouds have mostly dissipated on the night side (see discussion of \citealt{Roman2021} below, with notable caveats that \citealt{Helling2019} find nightside clouds in models of WASP-18b, which has an equilibrium temperature near 2400 K). \added{This dependence of temperature on phase amplitude is also discussed in \cite{Kataria2016}.}

\cite{Zhang2018} saw evidence in published data for a decrease in hot spot offset with increasing irradiation temperature, with an inflection point and increasing offsets after 3500 K, however this disagrees with some 3D model predictions such as \cite{Perna2012,Kataria2016,Komacek2017,Zhang2017,Parmentier2021}. \added{In fact, \cite{Kataria2016} predict the opposite, that the infrared phase offset will be smaller for the hottest planets that they modeled.} Observational results from \cite{Beatty2019} and \cite{Keating2019} also disagree with this offset trend while also suggesting that the night sides of all hot Jupiters are all roughly 1000 K due to clouds. \cite{Bell2021} similarly present little evidence for a trend between hot spot offset and temperature in the observed sample, but see evidence for the same increase in day-night contrast with temperature as previously seen in observations and predicted by models. 

More recent modeling by \cite{Roman2021} looks at bolometric phase-dependent emission and predict a dissipation of most nightside clouds near an equilibrium temperature of 2000K, with cloudy and clear phase curves converging past this point. For clear atmospheres, they predict a decrease of hot spot offset and increase in amplitude with increasing temperature. Exact trends for their cloudy models depends on assumptions made, with general peaks in phase offset around T$_{eq}\sim$ 2000K while amplitudes decrease then flatten off around the same temperature. \cite{Parmentier2021}, expanding on the work of \cite{Parmentier2016}, do not extend their models past T$_{eq}\sim$ 2000K, primarily focusing on the differences between clear and cloudy cases. Their clear models broadly agree with \cite{Roman2021} while the shape and inflection points of their relationships between offset/amplitude and equilibrium temperature in their cloudy models depends on the assumed cloud composition and particle size. \cite{Parmentier2021} also present model predictions for the apparent redistribution factor ($T_b^4/T_{eq}^4$, where $T_b$ refers to brightness temperature and $T_{eq}$ the equilibrium temperature) in the \spitzer\ band passes, for both cloudy and cloud-free GCMs, as a function of equilibrium temperature from 1000 - 2000 K.

In this work, we take steps towards completing the analysis of the remaining unpublished \spitzer\ phase curves and present results for seven new \spitzer\ phase curves from four planets, Qatar-2b, WASP-52b, WASP-34b, and WASP-140b, as well as a reanalysis of the 4.5 $\micron$ Qatar-1b phase curve using our fixed intrapixel sensitivity map. Five of these phase curves were observed at 4.5 $\micron$ and three at 3.6 $\micron$. We present as uniform of a data reduction as possible, with a consistent use of systematic models (Standard BLISS or fixed BLISS maps) throughout. Standard BLISS and fixed BLISS maps generally agree within uncertainties, but the use of the fixed BLISS map when appropriate eliminates other degeneracies between the astrophysical and systematic signals. While we do not apply the exact same systematic method to each data set, the analysis is performed by a single person applying the same criteria which results in a more uniform reduction than has generally been done in the current literature \citep[notable exceptions include][which takes a uniform analysis approach]{Bell2021}. \cite{Keating2020} adopted a similar approach to study trends for a set of 3 planets with equilibrium temperatures $\sim$ 1400 K planets.

This uniform approach better enables population level trends, which we examine at the end of this paper \citep[similar large uniform analyses to enable population studies have been done on \spitzer\ eclipses and transits, e.g.][]{Baxter2021,Mansfield2021}. All five planets in our sample have similar equilibrium temperatures near 1300 K, allowing us to specifically focus on trends in secondary parameters such as orbital period and gravity. Theory predicts that circulation and heat redistribution of tidally locked hot Jupiter atmospheres is governed by non-dimensional parameters that depend on, among other things, rotation rate (directly related to orbital period for tidally locked hot Jupiters, which is assumed here) and gravity -- suggesting that trends in phase curve parameters due to secondary parameters are expected. For example, the ratio between the wave propagation timescale and radiative timescale governs the day-night heat transport, and therefore determines the phase curve amplitude \citep{PerezBecker2013, Komacek2016}. This temperature range also results in a more direct comparison to model predictions because it is below the threshold where one needs to consider the effects of magnetohydrodynamics (MHD) and Hydrogen dissociation, commonly not included in 3D models.

In Section \ref{obs} we introduce each planet and overview the observations of their respective data sets. Section \ref{analysis} discusses the data reduction techniques, including systematic and astrophysical models. Section \ref{results} presents our phase curve results, with comparisons between planets discussed in Section \ref{comp}. Conclusions of this work are discussed in Section \ref{conclusions}. \added{Files containing phase curve data, model fits, and parameters used in our population studies are available \href{https://github.com/erinmmay/Spitzer_Uniform_Phase_Curves}{here}.}

\section{{\em Spitzer} Observations}
\label{obs}
\begin{figure*}
    \centering
    \includegraphics[width=0.24\textwidth]{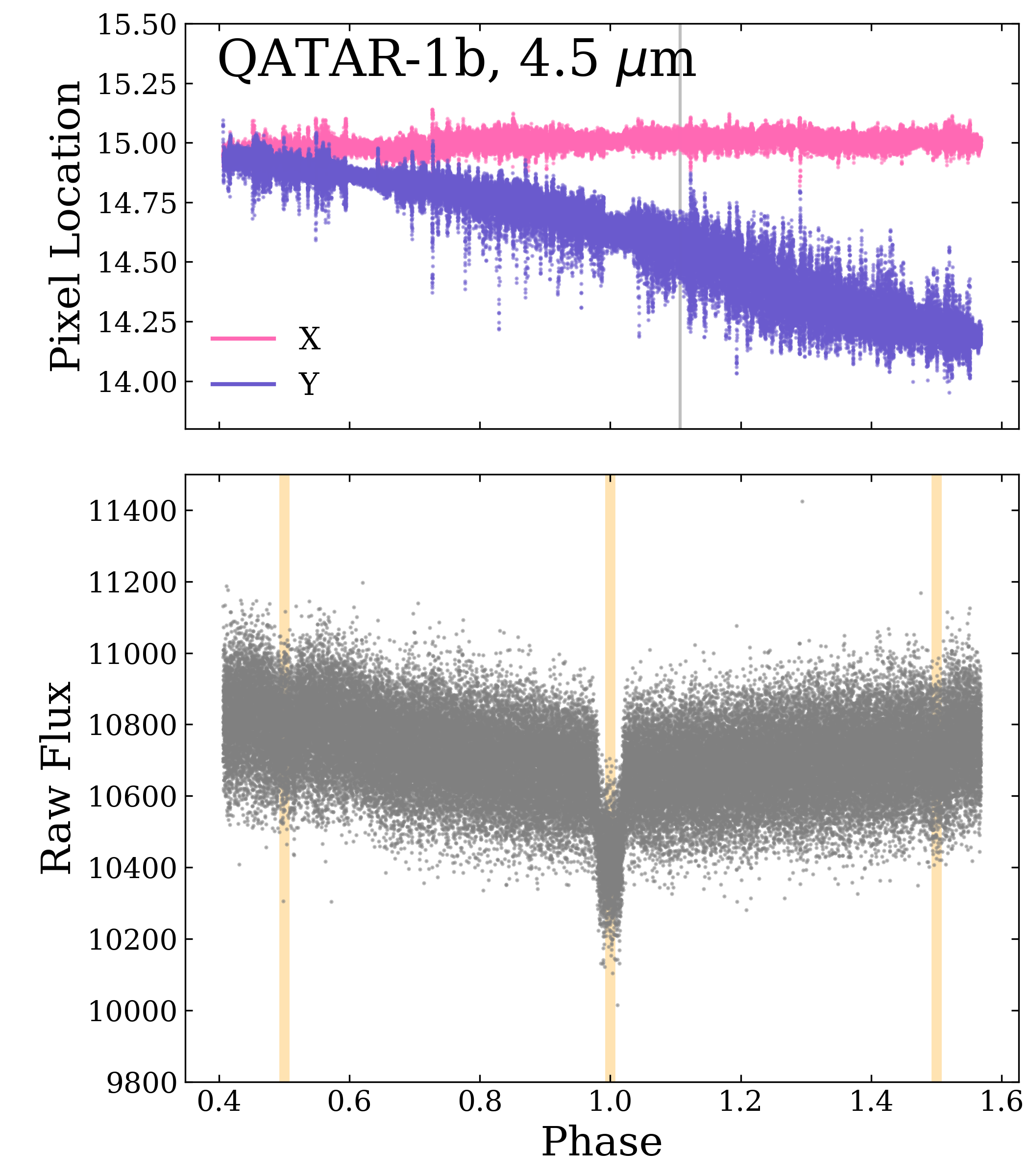}
    \includegraphics[width=0.24\textwidth]{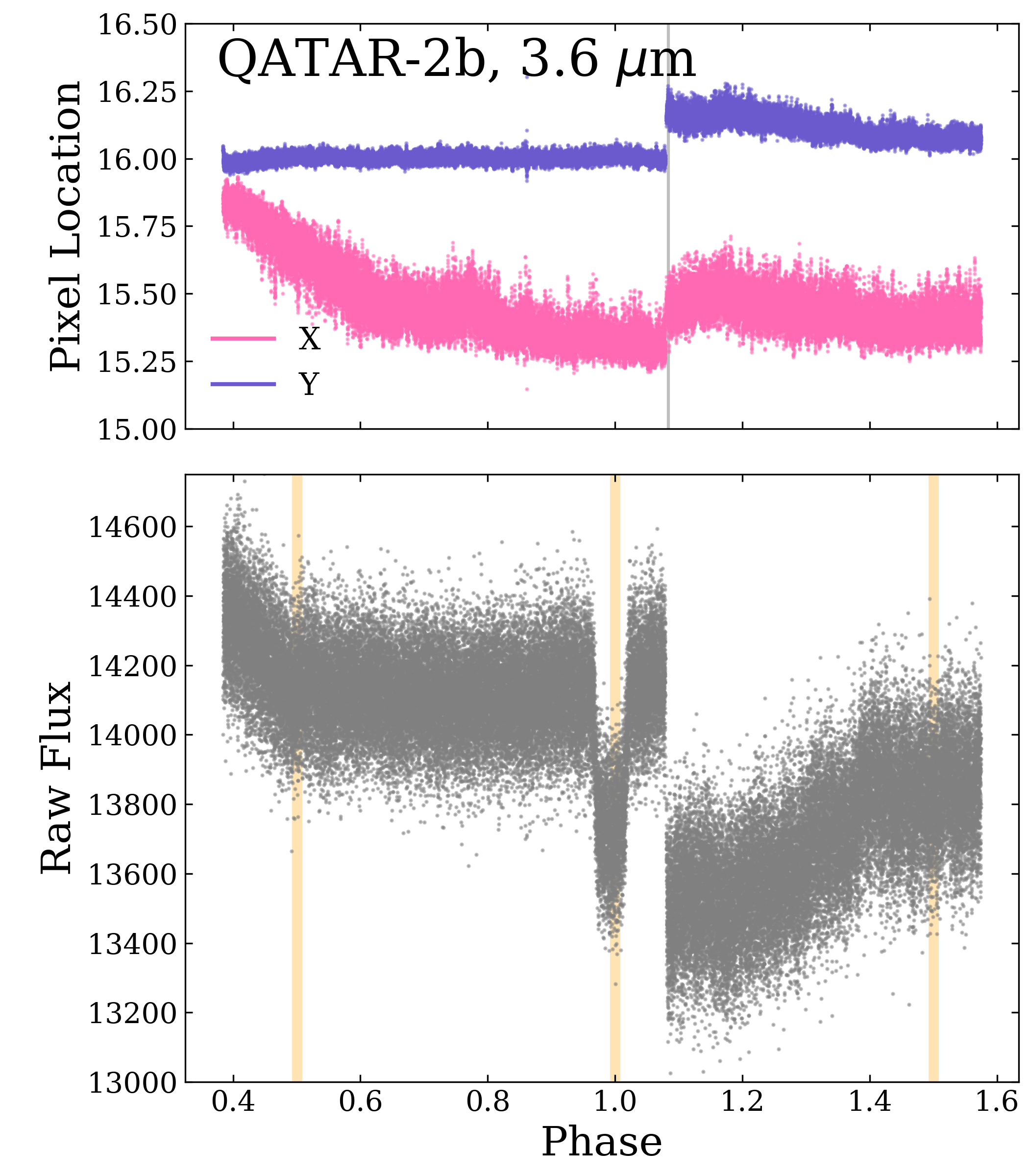}
    \includegraphics[width=0.24\textwidth]{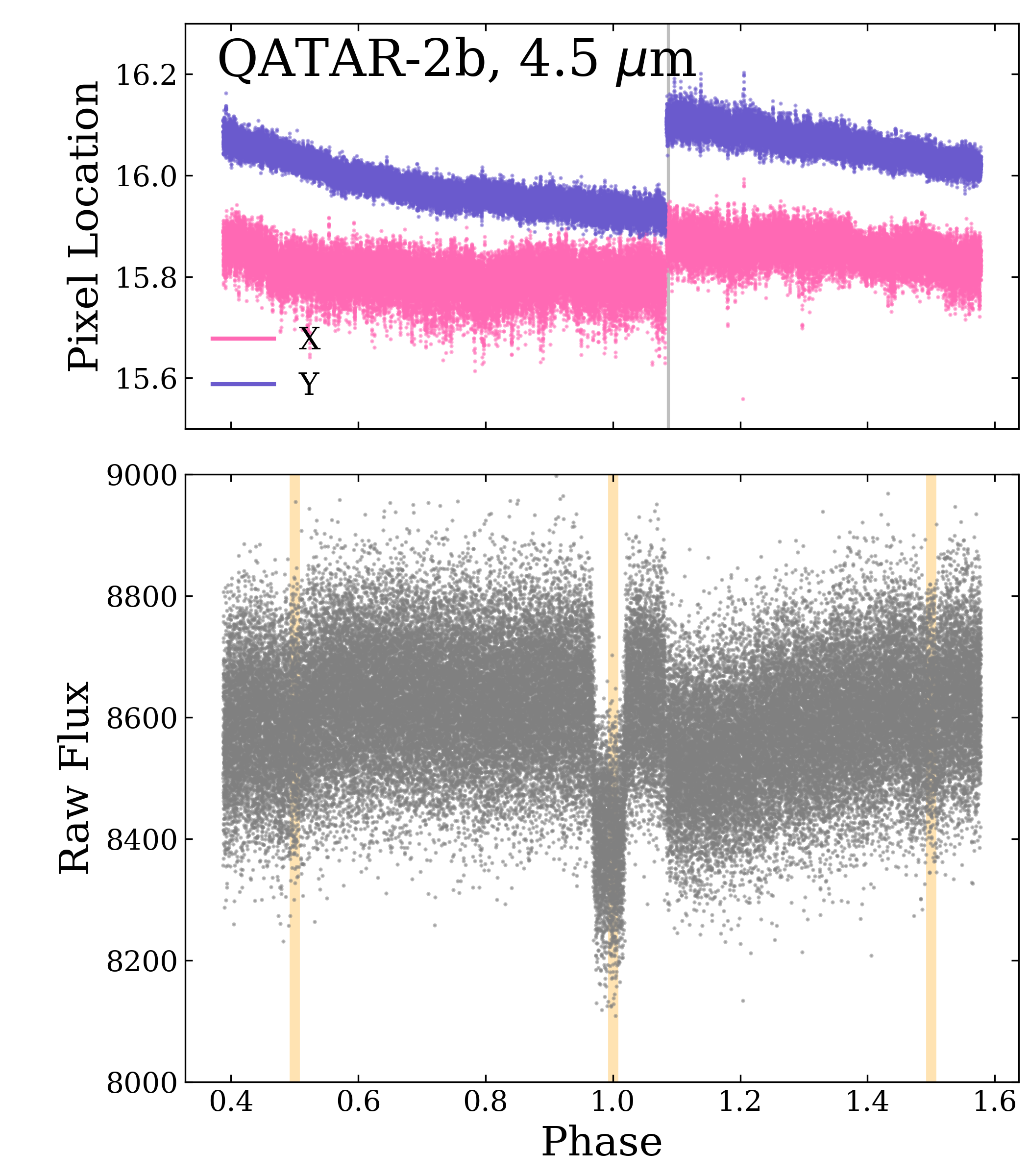}
    \includegraphics[width=0.24\textwidth]{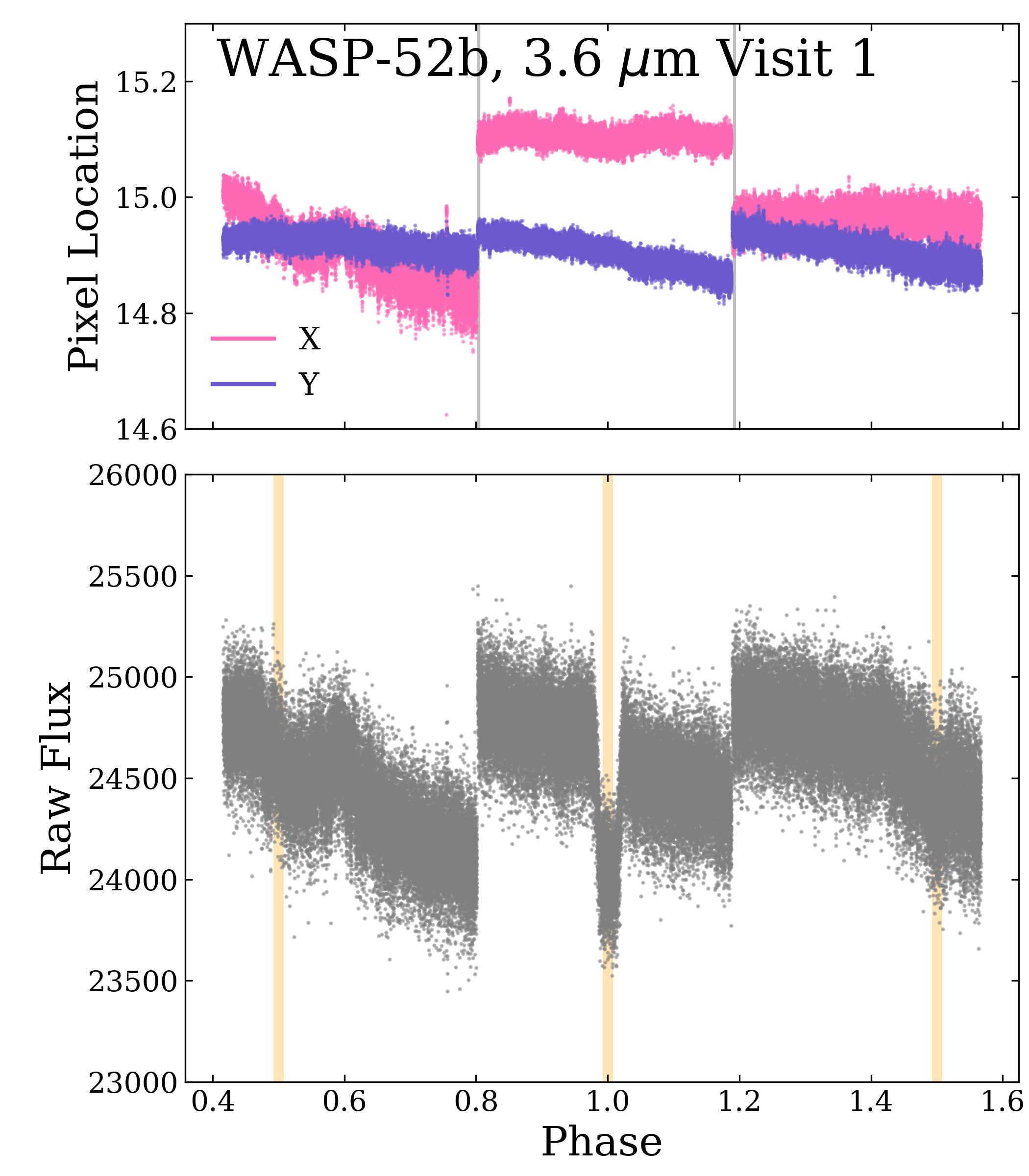}
    \includegraphics[width=0.24\textwidth]{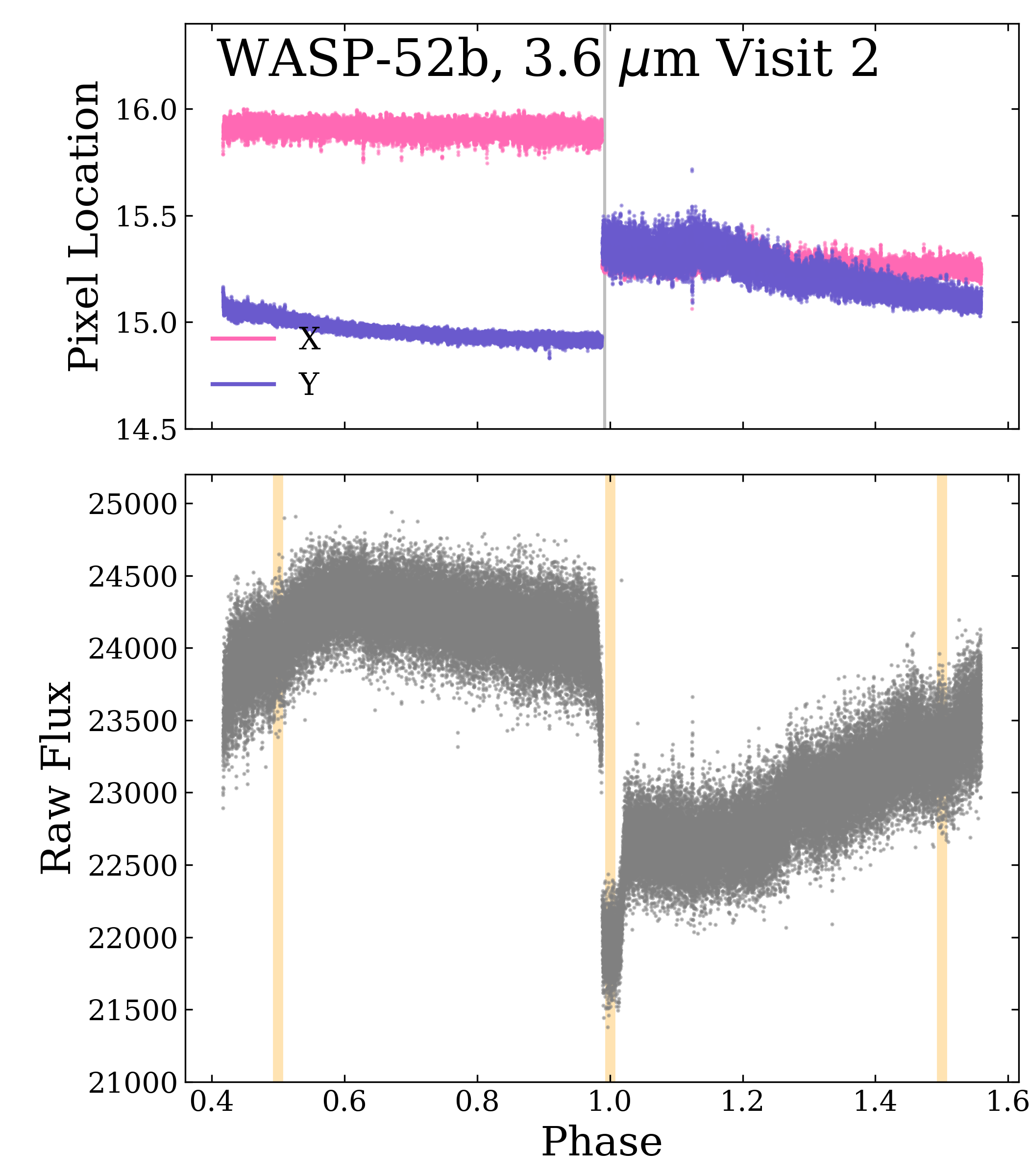}
    \includegraphics[width=0.24\textwidth]{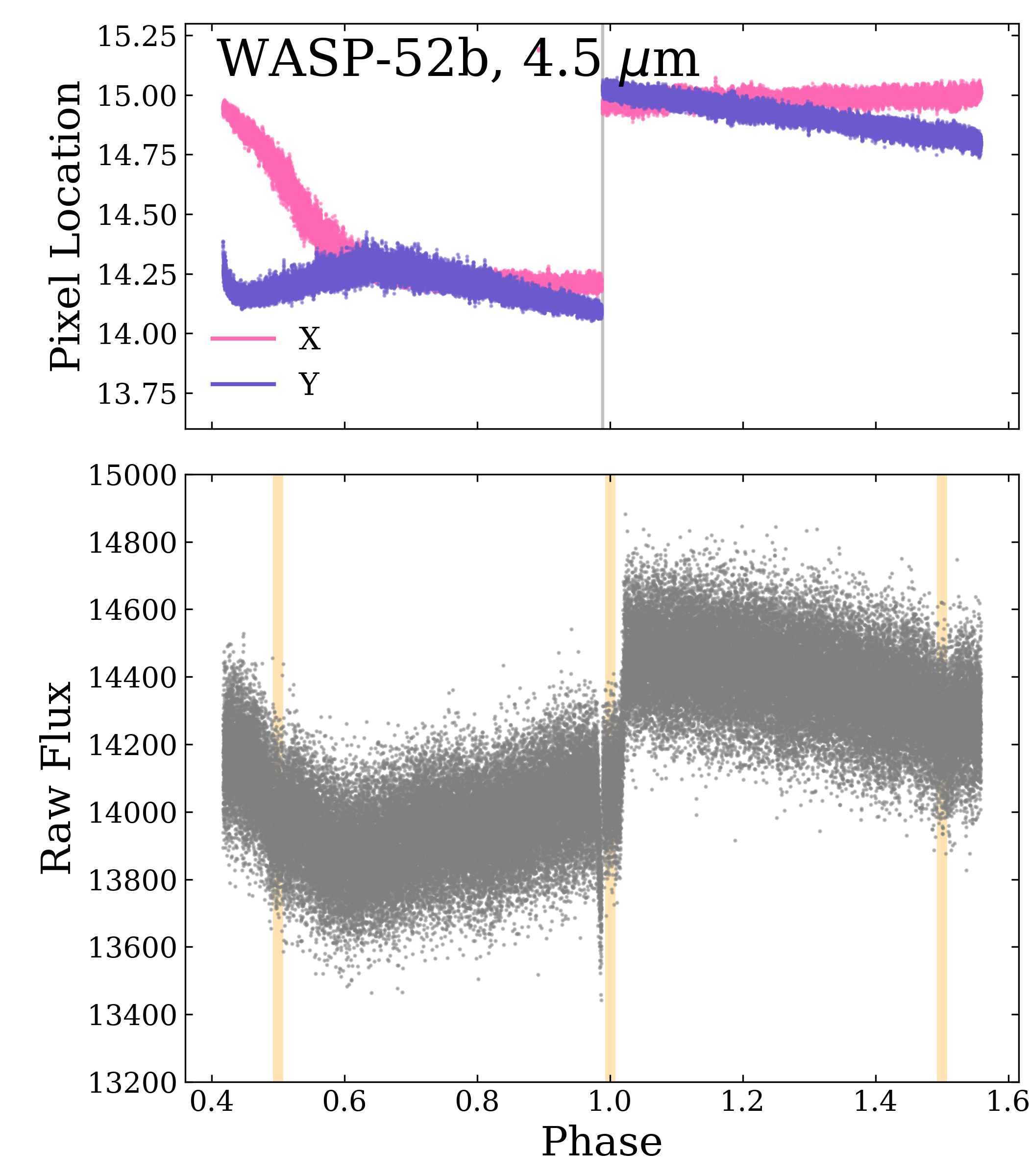}
    \includegraphics[width=0.24\textwidth]{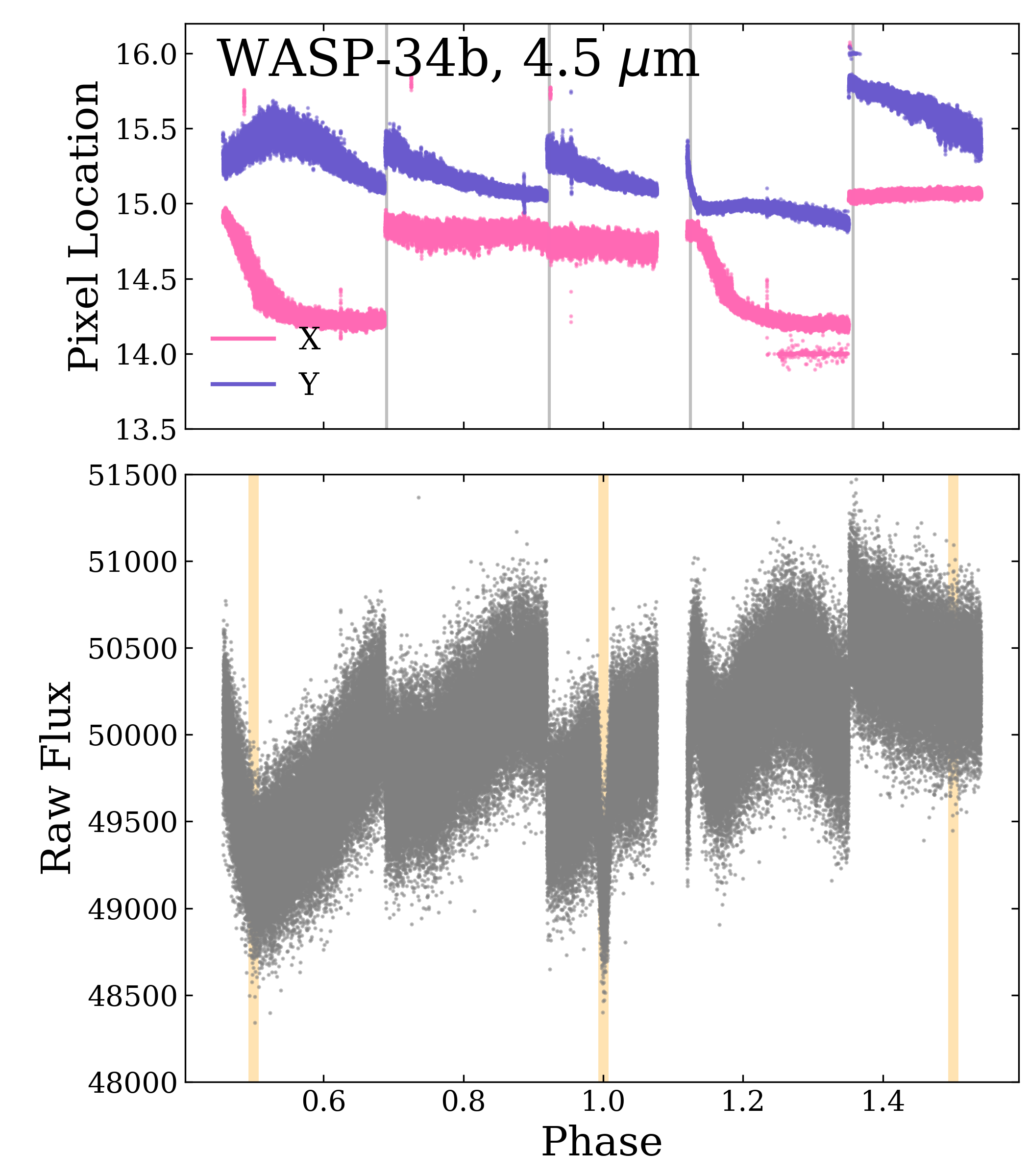}
    \includegraphics[width=0.24\textwidth]{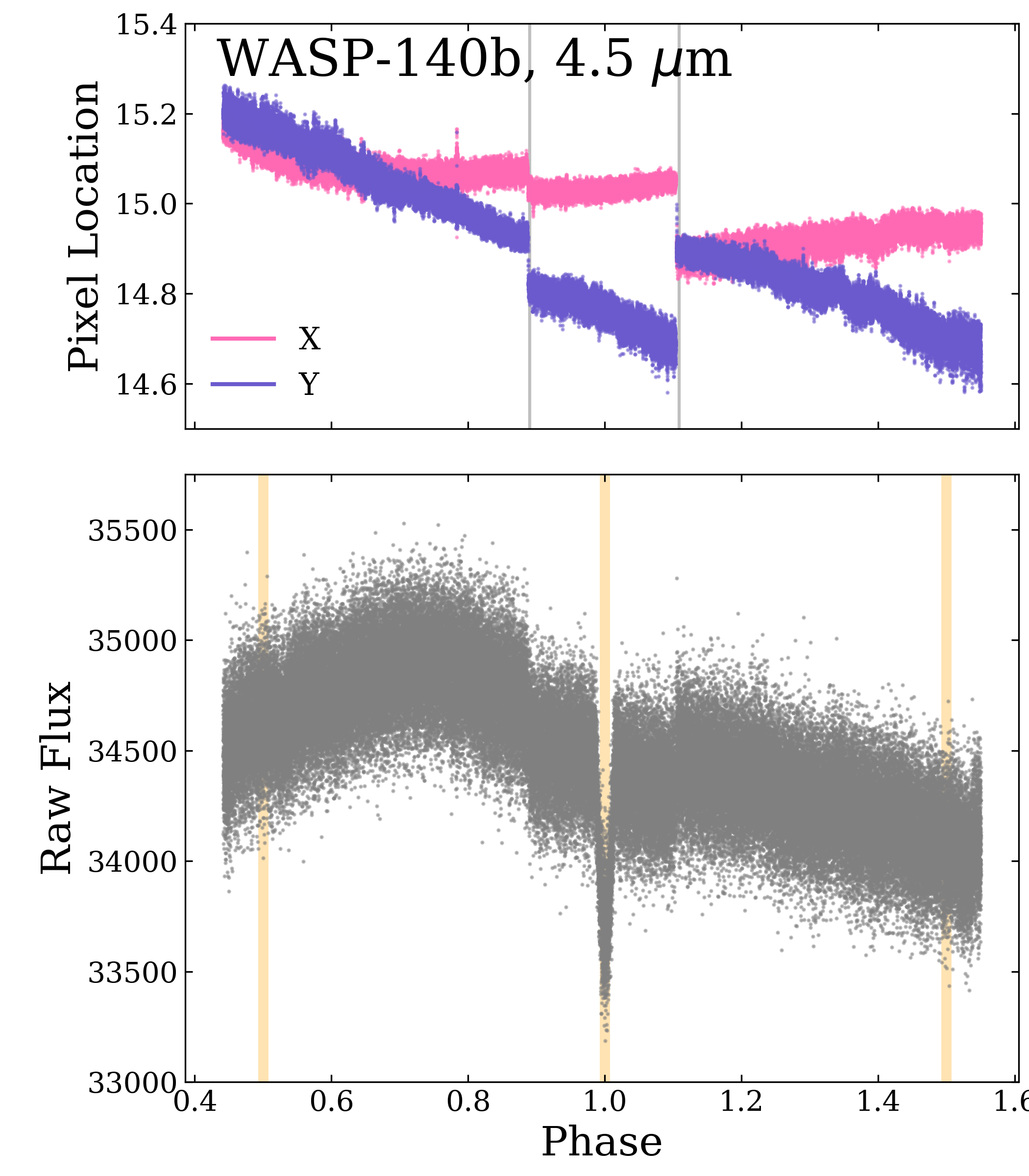}
    \caption{X and Y centroids (\textbf{top}) and corresponding raw (uncorrected) flux (\textbf{bottom}) for all phase curves analyzed in this study. In the top panel, the AOR gaps are marked and in the bottom panel phases of $\pm$ 0.5 and 1.0 are marked to guide the eye. Full-size versions of each image are available on our Uniform Phase Curve Repository.}
    \label{fig:centroids}
\end{figure*}

Here we discuss the data sets used in this work. Table \ref{table:observations} provides an overview of the relevant data sets and Table \ref{table:planet params} provides an overview of relevant planetary parameters. All phase curves were observed by \spitzer's InfraRed Array Camera \citep[IRAC,][]{Fazio2004}. As is standard, each phase curve starts before a secondary eclipse event and contains 2 secondary eclipse events and one transit event. Containing two eclipses allows for two points of reference that help minimize degeneracies between visit long systematic and astrophysical trends. Figure \ref{fig:centroids} shows the changes in $x$ and $y$ centroids over the course of all phase curve observations, as well as the corresponding raw flux. 

\begin{deluxetable*}{c c c c c c}
    \tablecolumns{6}
    \tabletypesize{\footnotesize}
    \tablecaption{Observational Details}
    \label{table:observations}
    \tablehead{
    \colhead{Label} &
        \colhead{Observation} &
        \colhead{Duration} \vspace{-0.2cm}&
        \colhead{Frame.} &
        \colhead{Total} &
        \colhead{Band} \\ 
          & \colhead{Date} & \colhead{(hrs)} & \colhead{Time (s)} & \colhead{Frames} & \colhead{($\micron$)}
                }
    \startdata
        \hline \hline
        qa001bo21   & May 02-03 2018  & 39.6    &  2.0  & 70,464    & 4.5   \\ \hline
        qa002bo11   & May 29-31 2017  & 38.2    &  2.0  & 67,904    & 3.6 \\ 
        qa002bo21   & May 21-23 2017  & 38.2    &  2.0  & 67,904    & 4.5   \\ \hline
        wa052bo11   & Oct 17-19 2016  & 48.3    &  2.0  & 85,824    & 3.6 \\ 
        wa052bo12   & Oct 21-23 2017  & 47.9    &  2.0  & 85,248    & 3.6 \\
        wa052bo21   & Oct 21-23 2018  & 47.9    &  2.0  & 85,248    & 4.5 \\ \hline
        wa034bo11   & Nov 03-08 2020  & 112.4   &  2.0  &  191,424  &   4.5 \\ \hline
        wa140bo11   & Jan 01-04 2019  & 59.5    &  2.0  &  105,728  &    4.5 \\ \hline
    \enddata
    \tablecomments{Label denotes the planet (e.g. qa001b = Qatar-1b), type of observation (o=orbit), \spitzer\ IRAC channel (1 or 2) and visit number (1 or 2)}
\end{deluxetable*}

\subsection{Qatar-1b}
One phase curve each at 3.6 $\micron$ and 4.5 $\micron$ were obtained as part of program 13038 (PI: Kevin Stevenson). Qatar-1b is a 1.294$^{+0.052}_{-0.029}$ M$_{\rm{Jup}}$, 1.143$^{+0.026}_{-0.025}$ R$_{\rm{Jup}}$ planet with an equilibrium temperature of 1360 $\pm$ 28 K \citep[using the most recent complete set of parameters from][]{Collins2017} and was first identified by \cite{Alsubai2011}. The phase curves from this program were previously analyzed and reported by \cite{Keating2020} and \cite{Bell2021}, here we reanalyze the 4.5 $\micron$ observation using the uniform \spitzer\ sensitivity map at 4.5 $\micron$ presented in \cite{May2020}. \spitzer\ eclipses of Qatar-1b (one each at 3.6 and 4.5 $\micron$) were analyzed by \cite{Garhart2018} and included in a statistical analysis of secondary eclipses by \cite{Garhart2020}.

\subsection{Qatar-2b}
One phase curve each at 3.6 $\micron$ and 4.5 $\micron$ of Qatar-2b were obtained as part of program 13038 (PI: Kevin Stevenson). Qatar-2b is a 2.487 $\pm$ 0.086 M$_{\rm{Jup}}$, 1.144 $\pm 0.035$ R$_{\rm{Jup}}$ planet with an equilibrium temperature of 1290 $\pm$ 14 K  \citep[using the most recent complete set of parameters from][]{Mocnik2017}. The discovery of Qatar-2b was reported by \cite{Bryan2012}.

\subsection{WASP-52b}
Two 3.6 $\micron$ and one 4.5 $\micron$ phase curves of WASP-52b were observed as a part of program 13038 (PI: Kevin Stevenson). WASP-52b is a 0.46 $\pm$ 0.02 M$_{\rm{Jup}}$, 1.27 $\pm$ 0.03 R$_{\rm{Jup}}$ planet with an equilibrium temperature of 1300 $\pm$ 35 K \citep[using the most recent complete set of parameters from ][]{Ozturk2019}. WASP-52b was first identified by \cite{Hebrard2013}. The transits from the first 3.6 $\micron$ visit and the single 4.5 $\micron$ visit were reduced and reported by \cite{Alam2018}.

\subsection{WASP-34b}
One 4.5 $\micron$ phase curve of WASP-34b was observed as part of program 14059 (PI: Jacob Bean). With an orbital period of 4.317 days, this was the longest \spitzer\ phase curve of a standard hot Jupiter observed during the telescope's lifetime \added{(HAT-P-2b, while on a longer orbit, is eccentric and not use in our comparison studies)}. WASP-34b is a 0.59 $\pm$ 0.01 M$_{\rm{Jup}}$, 1.22 $^{0.11}_{-0.08}$ R$_{\rm{Jup}}$ planet with an equilibrium temperature of 1158 $\pm$ 30 K \citep[using the most recent complete set of parameters from the discovery paper][]{Smalley2011}. Notably, WASP-34b has a grazing transit and eclipse, with an impact parameter of 0.904 $^{+0.017}_{-0.014}$.

\subsection{WASP-140b}
One 4.5 $\micron$ phase curve of WASP-140b was observed as part of program 14059 (PI: Jacob Bean). WASP-140b is a 2.44 $\pm$ 0.07 M$_{\rm{Jup}}$, 1.44 $^{+0.42}_{-0.18}$ R$_{\rm{Jup}}$ planet with an equilibrium temperature of 1320 K \citep{Hellier2017}. WASP-140b is the only eccentric planet in this study, with an eccentricity of 0.0470 $\pm$ 0.0035. The star likely is relatively young at 0.42 $\pm$ 0.06 Gyr (\citealt{Hellier2017} based on a 10.4 rotational period and ages from \citealt{Barnes2007}), which, compared to the circularization timescale of the orbit, suggests that WASP-140b only recently arrived at its present location. Although WASP-140b does have the same equilibrium temperature as the rest in this sample, the eccentricity puts WASP-140b in a category on its own. Regardless, we include the analysis here for completeness.

\begin{deluxetable*}{c c c c c c c l }
    \tablecolumns{8}
    \tabletypesize{\footnotesize}
    \tablecaption{Planet Parameters}
    \label{table:planet params}
    \tablehead{
    \colhead{Planet} &
        \colhead{a} &
        \colhead{log(g)} &
        \colhead{Radius} &
        \colhead{Mass} &
        \colhead{T$_{eq}$} &
        \colhead{Period} &
        \colhead{Reference} \\
         {} & (AU) & (cm/s$^2$) & (R$_{\rm{Jup}}$) & (M$_{\rm{Jup}}$) & (K) & (days)  &  }
    \startdata
        \hline \hline
        Qatar-1b   & 0.0233 $\pm$ 0.0040  & 3.390 $\pm$ 0.015 &  1.143$^{+0.026}_{-0.025}$ & 1.294$^{+0.052}_{-0.049}$ & 1360 $\pm$ 28 & 1.42002420 $\pm$ 2.2E-7  & \cite{Collins2017}   \\ \hline
        Qatar-2b  &  0.02149 $\pm$ 0.00036 & 3.638 $\pm$ 0.022 & 1.144 $\pm$ 0.035 & 2.487 $\pm$ 0.086 & 1290 $\pm$ 14 & 1.3371182 $\pm$ 3.7E-6  &  \cite{Bryan2012}   \\ \hline
        WASP-52b  &  0.0272 $\pm$ 0.003 &  2.81 $\pm$ 0.03  & 1.27 $\pm$ 0.03 & 0.46 $\pm$ 0.02 & 1300 $\pm$ 35 & 1.7497798 $\pm$ 1.2E-6  & \cite{Hebrard2013} \\ \hline
        WASP-34b   & 0.0524 $\pm$ 0.0004 & 2.96 $^{+0.05}_{-0.07}$ & 1.22$^{+0.11}_{-0.08}$ & 0.59 $\pm$ 0.01 & 1158 $\pm$ 30 & 4.3176782 $\pm$ 4.5E-6  & \cite{Smalley2011}   \\ \hline
        WASP-140b  & 0.0323 $\pm$ 0.0005 & 3.4 $\pm$ 0.2 & 1.44$^{+0.42}_{-0.18}$ & 2.44 $\pm$ 0.07 & 1317 $\pm$ 40 & 2.2359835 $\pm$ 8E-7   & \cite{Hellier2017}   \\  \hline
    \enddata
    \tablecomments{Equilibrium temperatures are calculated from values in this table and Table \ref{table:LimbDarkening}.}
\end{deluxetable*}
%

\section{Data Reduction and Analysis} \label{analysis}
\subsection{Initial Data Reduction}
Data reduction and analysis is done with the Photometry for Orbits, Eclipses, and Transits \citep[POET,][]{Campo2011,Stevenson2012, Cubillos2013} pipeline, including recent updates from \cite{May2020} to improve systematic modeling at 4.5 $\micron$ by applying a fixed sensitivity map rather than self-calibrating. We use 2D Gaussian centroiding following the suggestions of \cite{Lust2014}. All data sets are extracted using a fixed aperture size optimized for the standard deviation of the normalized residuals (SDNR). To determine the best aperture, we extract apertures between 2.0 and 4.0 pixels in 0.25 pixel increments. For all data sets we use a fixed annulus between 7 and 15 pixels away from the centroids for background subtractions. The best aperture size for each data set is included in Table \ref{table:bestfits}.

\subsubsection{The Intrapixel Sensitivity Effect}
The dominant sources of \spitzer\ IRAC systematics at 3.6 and 4.5 $\micron$ are intrapixel sensitivity variations as the centroid drifts within a single pixel. We use Bilinearly Interpolated Subpixel Sensitivity (BLISS) mapping \citep{Stevenson2012} to model and remove this effect. In \cite{May2020} we showed that BLISS mapping is degenerate with point response function at full width half max (PRF FWHM) detrending. PRF-FWHM detrending is a second level detrending function that accounts for the shape of the PRF stretching towards an oval for centroids near the edges of the pixel (see Section \ref{PRF}). When temporally binning the data, this degeneracy becomes stronger due to smoothing over the features that allow you to fit the two methods independently. Therefore, we do not perform any temporal binning on any of the data sets presented in this work. 

\begin{figure*}
    \centering
    \includegraphics[width=0.24\textwidth,trim={0.75cm 0.75cm 1cm 1cm},clip]{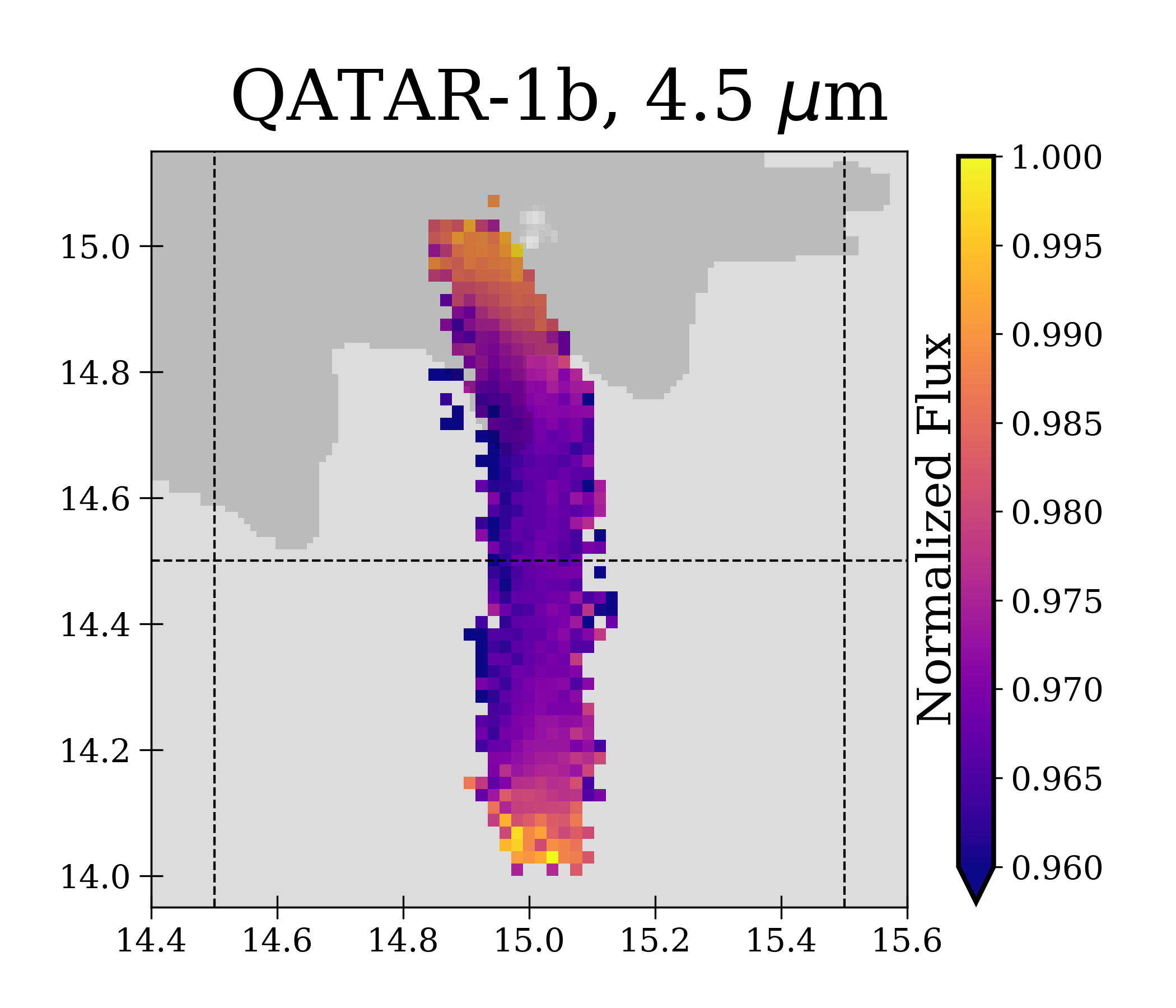}
    \includegraphics[width=0.24\textwidth,trim={0.75cm 0.75cm 1cm 1cm},clip]{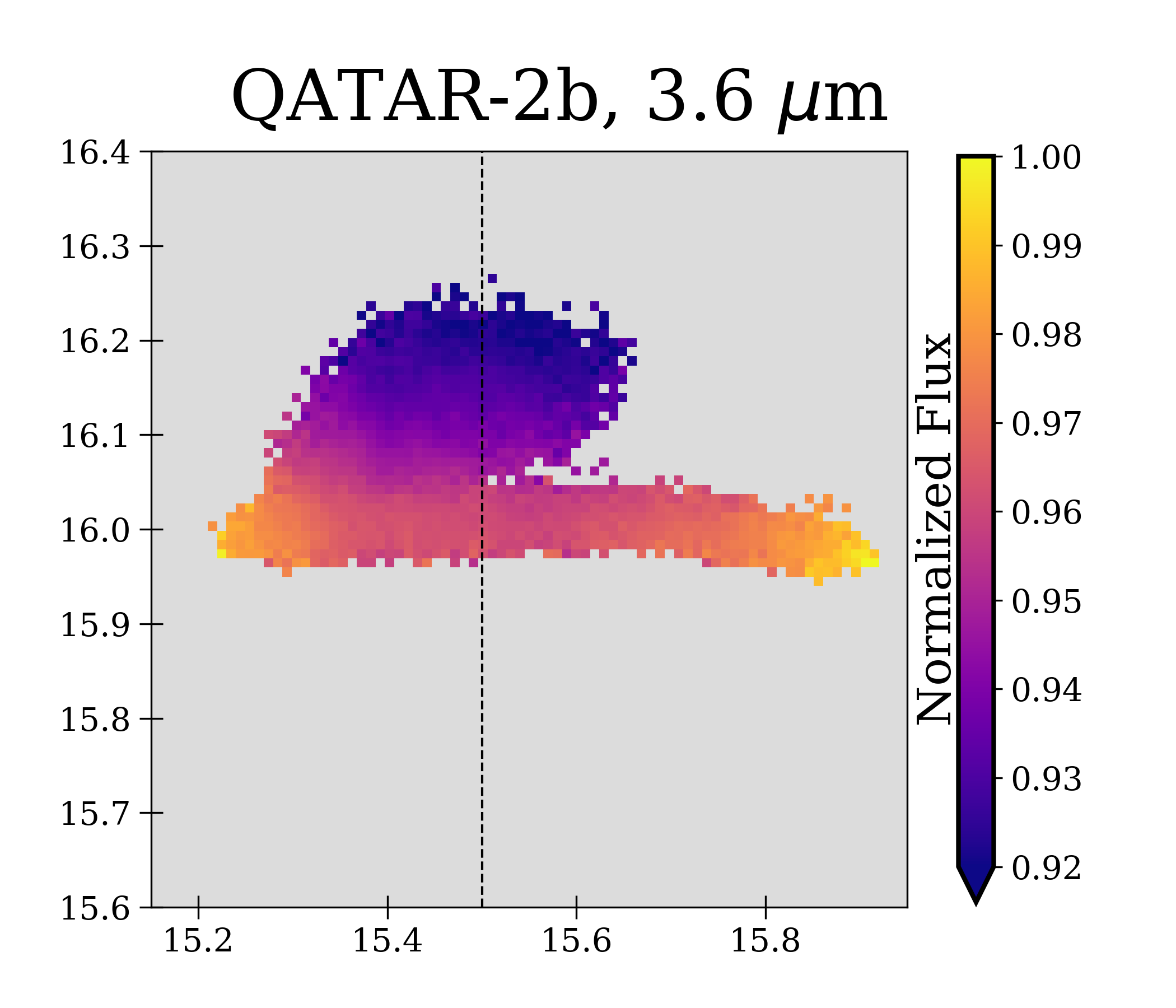}
    \includegraphics[width=0.24\textwidth,trim={0.75cm 0.75cm 1cm 1cm},clip]{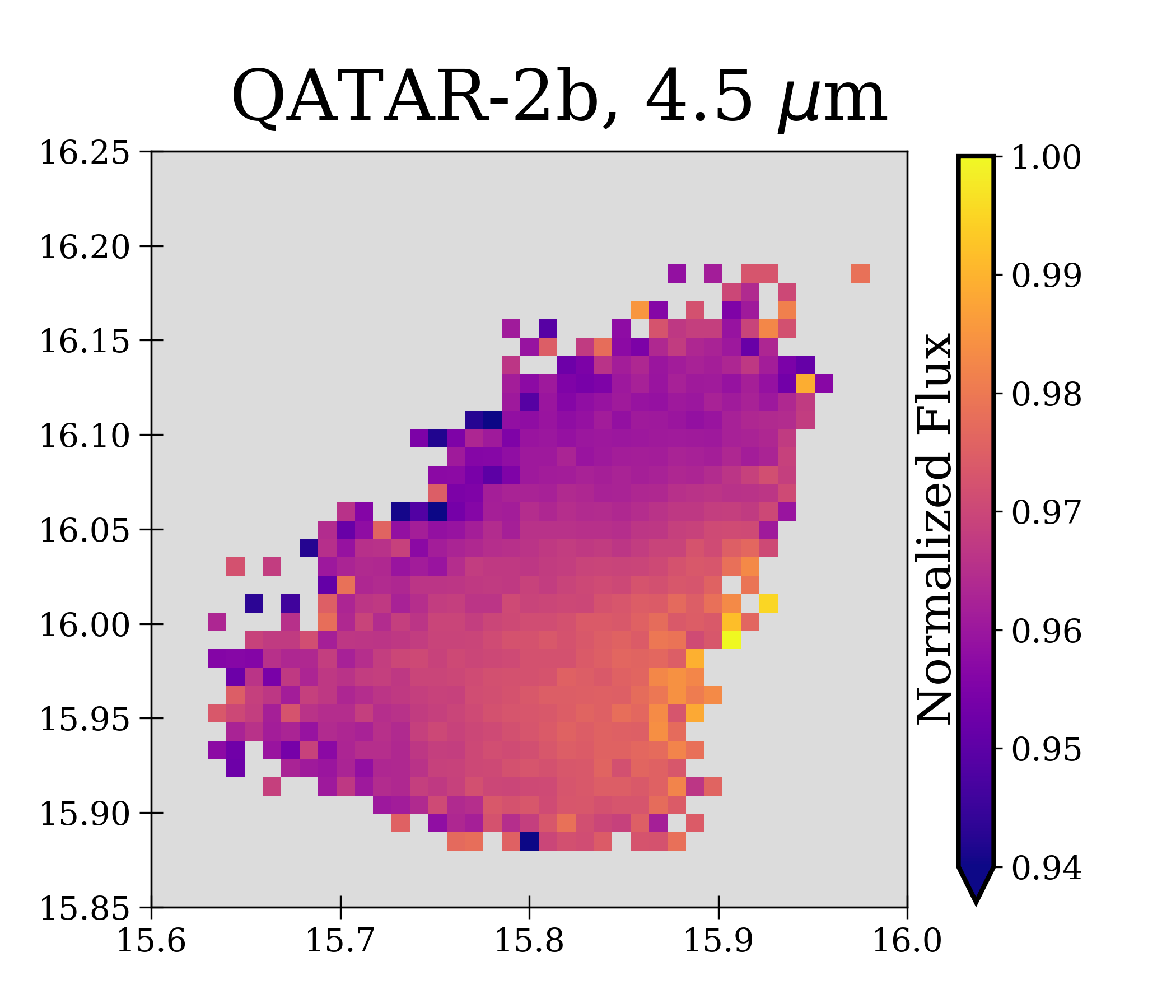}
    \includegraphics[width=0.24\textwidth,trim={0.75cm 0.75cm 1cm 1cm},clip]{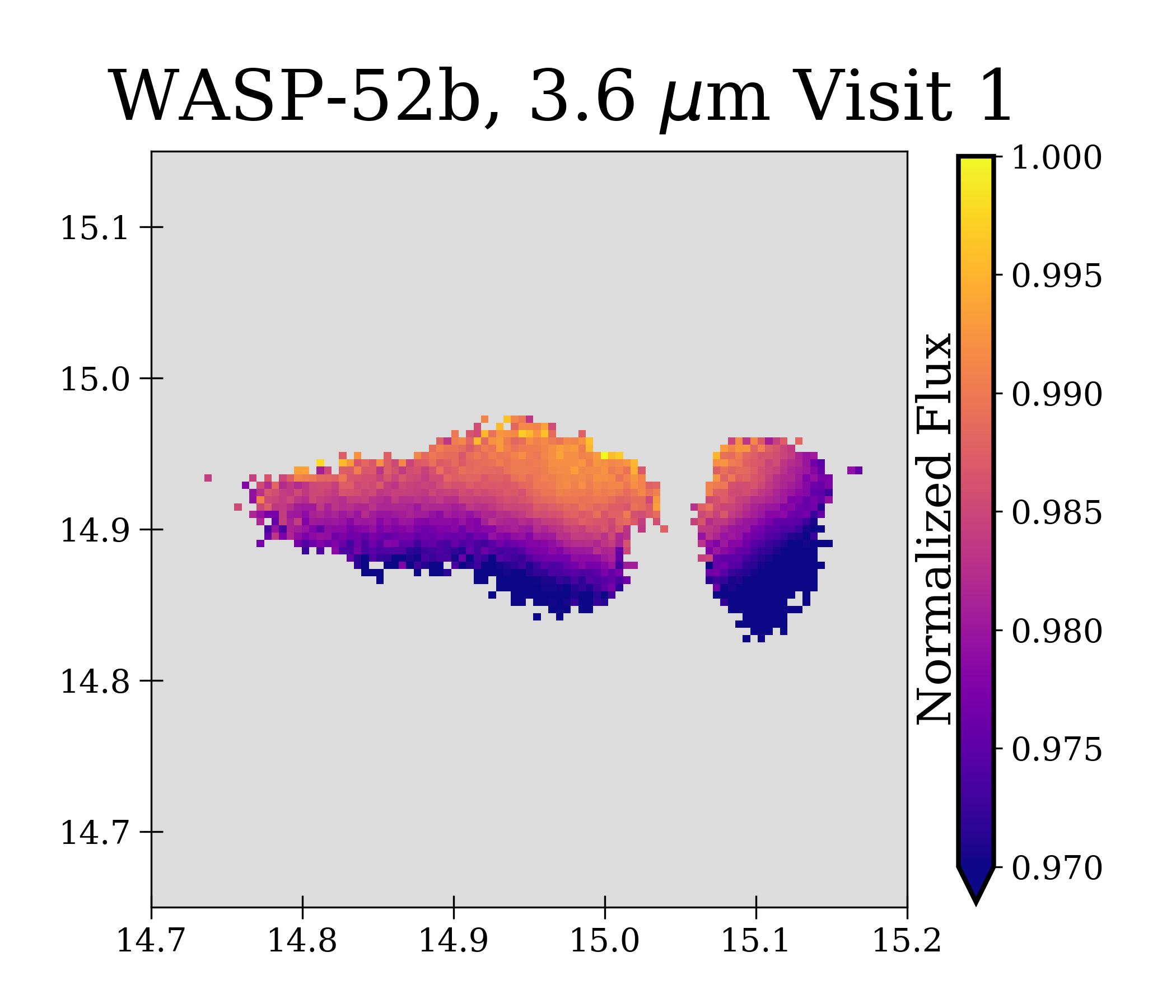}
    \includegraphics[width=0.24\textwidth,trim={0.75cm 0.75cm 1cm 1cm},clip]{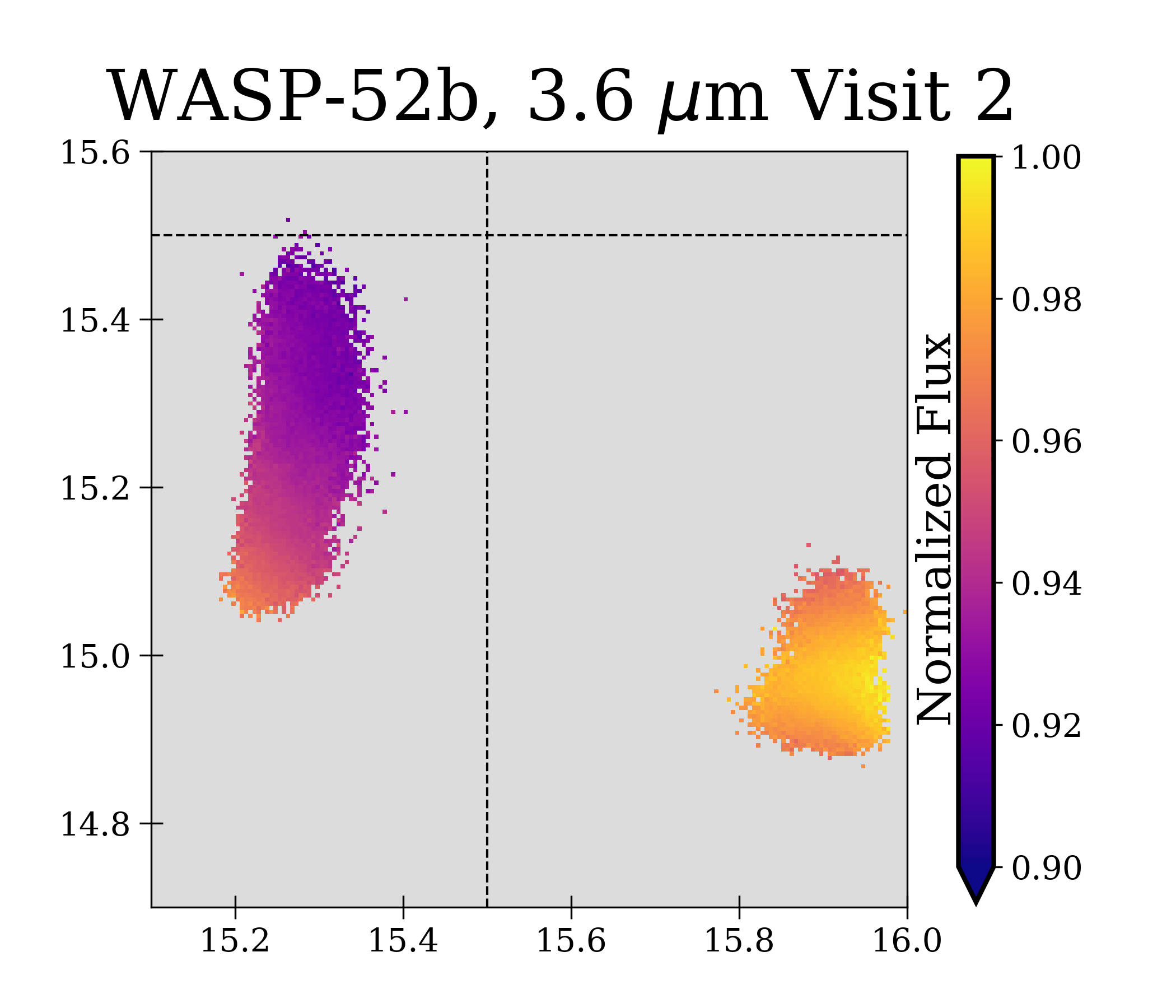}
    \includegraphics[width=0.24\textwidth,trim={0.75cm 0.75cm 1cm 1cm},clip]{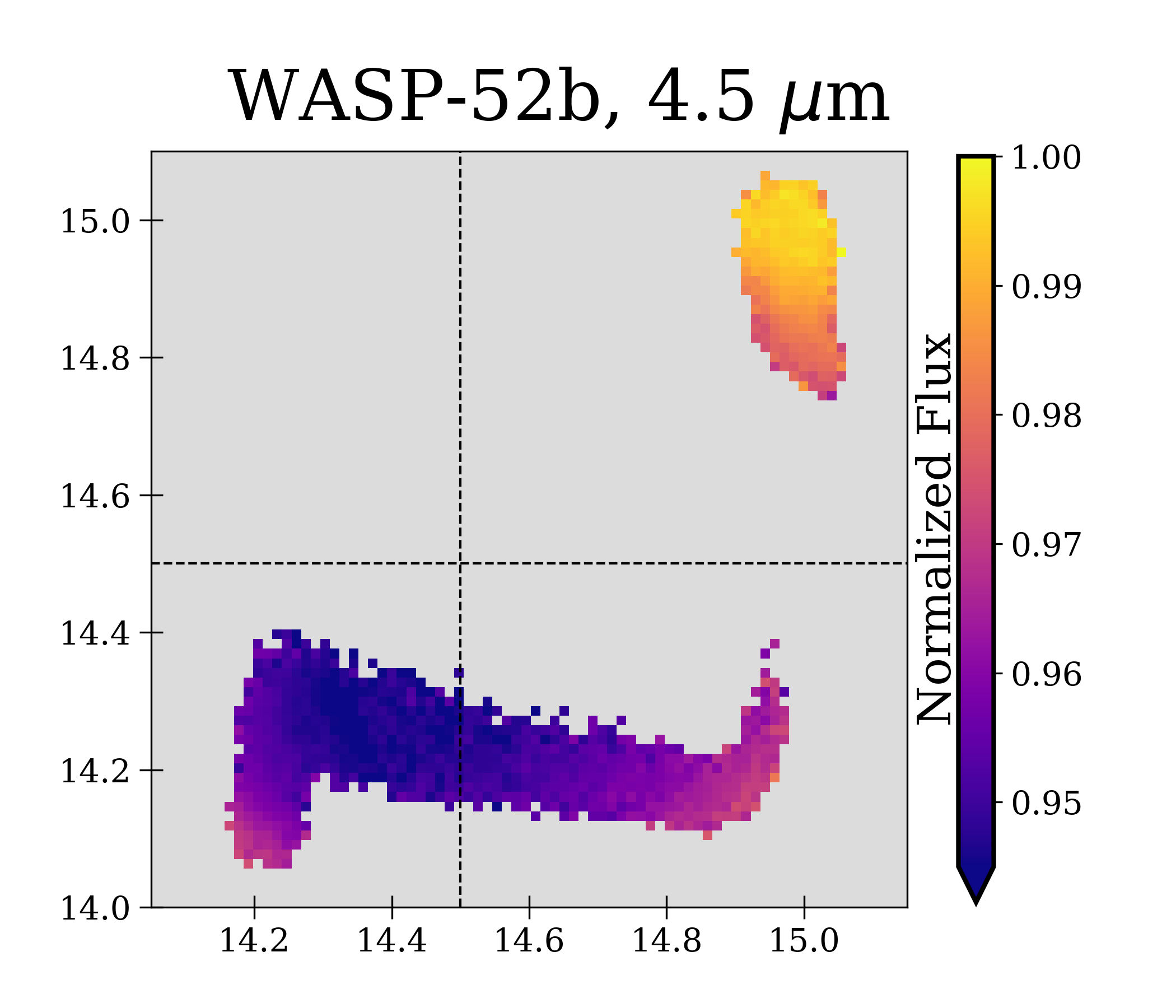}
    \includegraphics[width=0.24\textwidth,trim={0.75cm 0.75cm 1cm 1cm},clip]{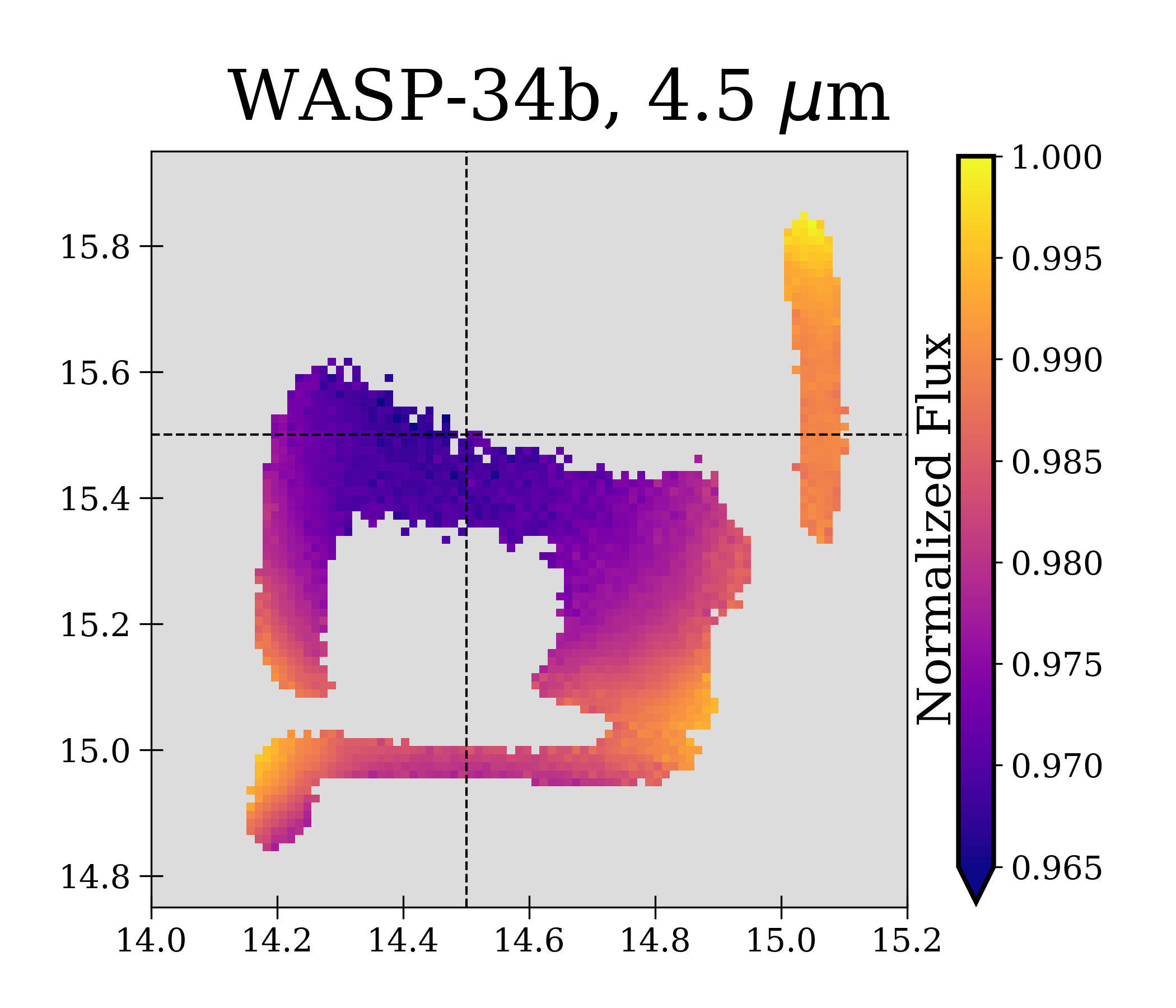}
    \includegraphics[width=0.24\textwidth,trim={0.75cm 0.75cm 1cm 1cm},clip]{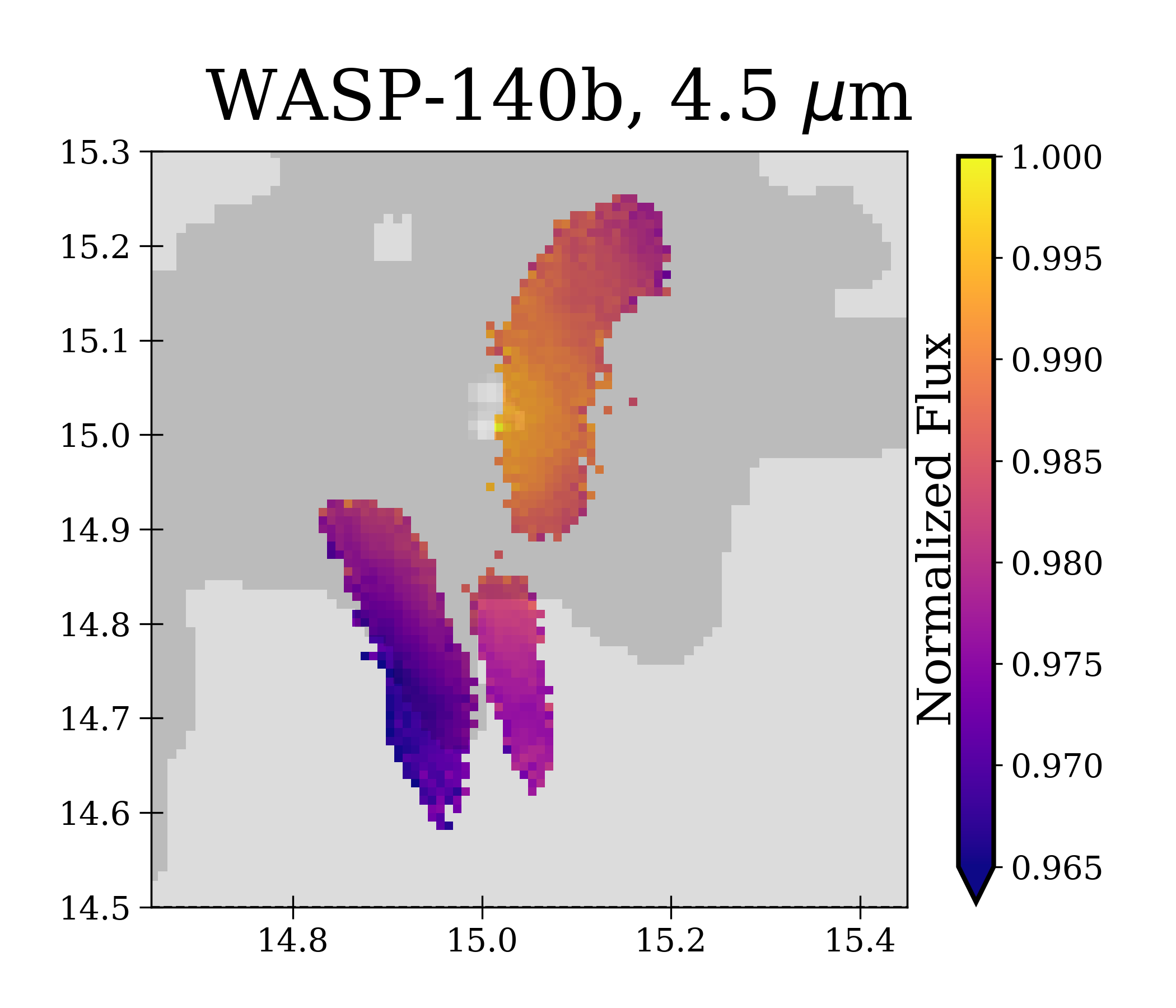}
    \caption{BLISS maps for the data sets analyzed in this work. The colorbar denotes the relative sensitivity of a given subpixel element. The 4.5 $\micron$ observations also show the extent of our fixed sensitivity map in the dark shaded region. The dashed lines denote the edges of a pixel (where relevant), with axis labels in sub-pixel units. Full-size versions of each image are available on our Uniform Phase Curve Repository.}
    \label{fig:BLISSmaps}
\end{figure*}

For the 4.5 $\micron$ data sets that fully or partially overlap with our fixed intrapixel sensitivity map, our systematic model uses the overlapping regions to detrend the data, and self-calibrates using standard BLISS mapping in non-overlapping regions. The transition from self-calibration to fixed map calibaration is smooth, with no artifacts. For the remaining 4.5 $\micron$ data sets that do not at least partially overlap with our fixed sensitivity map, the data sets fully self-calibrate using standard BLISS mapping. As discussed in \cite{May2020}, 3.6 $\micron$ sensitivity is time variable and a fixed sensitivity map cannot be generated. As a result, all 3.6 $\micron$ data sets are self-calibrated with standard BLISS mapping. Figure \ref{fig:BLISSmaps} shows the best fit BLISS maps for all data sets in this study.

The standard BLISS map is described by the intrapixel spatial binning size and the minimum number of exposures required for a given spatial bin to be used in the fit. To determine the best set of these parameters we compare model fits to a nearest neighbor approach and identify the step size that produces the best BIC and SDNR, respectively, without overfitting the data. See \cite{Stevenson2012} for more details on our methods.  

\subsubsection{PRF Detrending} \label{PRF}
The IRAC point-response function (PRF) tends to stretch towards an oval rather than a circle as a centroid drifts towards the edge of a pixel. Because we use circular apertures, this can result in flux being under-counted. Previous works have used a method to correct this by detrending against the Gaussian widths of the PRF with varying polynomial orders \citep{Knutson2012, Lewis2013, Lanotte2014,Demory2016a,Demory2016b,Gillon2017,Mendonca2018} given as 
\begin{multline}
   f = x_1(s_x-s_0) + x_2(s_x-s_0)^2 + x_3(s_x-s_0)^3 +
    \\ y_1(s_y-s_0) + y_2(s_y-s_0)^2 + y_3(s_y-s_0)^3 + c 
\end{multline}
where $s_x$, $s_y$ are the x- and y- dimension Gaussian width in pixels, $s_0$ is an offset (typically 0.5 pixels), x$_{\{1,2,3\}}$ and y$_{\{1,2,3\}}$ are the polynomial coefficients and c is a constant. 

We test applying the PRF detrending when a free (i.e. standard) BLISS map is used, comparing 1$^{\rm{st}}$, 2$^{\rm{nd}}$, and 3$^{\rm{rd}}$ order polynomials and selecting for the lowest BIC. The fixed sensitivity map encapsulates the loss of flux from changing PRF shape, and does not require this step. In general, most of our data sets with the free map do prefer the addition of the PRF detrending with one exception, see Table \ref{table:bestfits}.

\subsubsection{Astrophysical Source Models}
We adopt a generic sinusoidal function to model each planet's emission considering both a full-period and half-period component \citep[][find that a sinusoid is the best model for fitting full phase curves]{Cowan2008}. The addition of a half-period sinusoidal function results in asymmetric phase curves. We use the {\tt{BATMAN}} package to model the transit events \citep{Kreidberg2015} with eclipses modeled using a form of the analytic method presented by \cite{Mandel2002}, with the exception of WASP-34b which is a grazing event. For the WASP-34b eclipses we used {\tt{BATMAN}} which includes an impact parameter input. All transits are modeled assuming quadratic limb darkening based on interpolated Kurucz stellar models \citep{Kurucz2004} using {\tt{ExoCTK}}'s limb darkening tool \citep{exoctk}. Table \ref{table:LimbDarkening} overviews the stellar parameters and limb darkening coefficients adopted for each host star. 

We further consider the presence of temporal ramps; including no-ramp and linear ramps, as well as quadratic or exponential ramp models for select data sets. The best fit combination of the systematic model, temporal ramp, and astrophysical sources is identified using the Bayesian Information Criterion \citep[BIC,][]{Liddle2008}, with the exception of model combinations that produce significantly non-physical events (i.e. relative flux ratios less than 1.0 for non-grazing events). We compute the best-fit models using a Levenberg-Marquardt minimizer, and our parameter uncertainties are estimated using a custom Differential-Evolution Markov Chain algorithm \citep[DEMC,][]{terBraak2008}. In Table \ref{table:bestfits} we present the $\Delta$BIC comparison of our various model combinations for all phase curves. The combination with the best BIC for each observation (shown as the bolded row) is identified as our `best-fit' model. In several cases the use of a quadratic ramp results in a nightside flux ratio less than 1.0, which we discard as nonphysical. \added{See Section \ref{appendix} for details about which paramters are fit and which are held constant. Each parameter takes a bounded uniform prior (e.g. time parameters are bounded based on the start and end times of the entire data set).}

In the following sections we discuss the results for each phase curve. 

\begin{deluxetable*}{c c c l l c c}
    \tablecolumns{6}
    \tabletypesize{\footnotesize}
    \tablecaption{Stellar Parameters and Limb Darkening}
    \label{table:LimbDarkening}
    \tablehead{
    \colhead{Star} &
        \colhead{T$_{eff}$} &
        \colhead{log(\textit{g})} &
        \colhead{[Fe/H]} &
        \colhead{Ch.} &
        \colhead{C$_1$} &
        \colhead{C$_2$} \\
        & (K) & (cm/s$^2$) & (dex) & ($\micron$) & & }
    \startdata
        \hline \hline
        Qatar-1   & 5013 $^{+93}_{-88}$ & 4.552 $^{+0.012}_{-0.011}$ & 0.17 $^{+0.097}_{-0.094}$ & 4.5   & 0.100  & 0.110 \\ \hline
        Qatar-2   & 4654 $\pm$ 50 & 4.601 $\pm$ 0.018 & 0.02 $\pm$ 0.08 & 3.6   & 0.116  & 0.166 \\ 
                   &  &  &                          & 4.5   & 0.107  & 0.120 \\ \hline
        WASP-52   & 5000 $\pm$ 100 & 4.582 $\pm$ 0.014 & 0.03 $\pm$ 0.12 & 3.6   & 0.108  & 0.147 \\ 
                   &  &  &                          & 4.5   & 0.100  & 0.107 \\ \hline   
        WASP-34   & 5700 $\pm$ 100 & 4.50 $\pm$ 0.10 & 0.040 $\pm$ 0.100$^{\dagger}$  & 4.5   & 0.090  & 0.099 \\ \hline
        WASP-140  & 5260 $\pm$ 100 & 4.51 $\pm$ 0.04 & 0.12 $\pm$ 0.10           & 4.5 & 0.095 & 0.105 \\ \hline
    \enddata
    \tablecomments{C$_1$ and C$_2$ are the quadratic limb darkening parameters. Unless otherwise noted stellar parameters are from the same source as planetary parameters in Table \ref{table:planet params}.
    \\ $^{\dagger}$ WASP-34b metallicity from \cite{Stassun2018}}
\end{deluxetable*}

\begin{deluxetable}{c c c | l l l}
    \tablecolumns{6}
    \tabletypesize{\scriptsize}
    \tablecaption{Best Fit Models}
    \label{table:bestfits}
    \tablehead{
    \colhead{Label} &
        \colhead{Aperture} & 
        \colhead{Systematic} \vspace{-0.2cm}&
        \colhead{Ramp} &
        \colhead{Phase} &
        \colhead{$\Delta$BIC} \\ 
         & \colhead{[Pixels]} & \colhead{Model} & \colhead{Model}  & \colhead{Model} &
        }
    \startdata
        \hline \hline
        qa001bo21   & 2.25  & Fixed &  --       & Symm. & 108.4\\
                    &       & BLISS &  Lin.     & Symm. & 5.9\\ 
                    &       &       &  \textbf{Quad.}  & \textbf{Symm.} & \textbf{0.0} \\ 
                    &       &       &  --       & Asymm. & 130.6\\ 
                    &       &       &  Lin.     & Asymm. & 28.2\\ 
                    &       &       &  Quad.    & Asymm. & 20.4\\ \hline \hline
        qa002bo11   & 2.00  & Free     &  --       & Symm. & 40.6 \\
                    &   & BLISS         &  Lin.     & Symm. & 5.4  \\ 
                    &   &  +            &  \textbf{Quad.}    & \textbf{Symm.} & \textbf{0.0} \\ 
                    &   & 2$^{nd}$      &  --       & Asymm. & 59.6 \\ 
                    &   & order         &  Lin.     & Asymm. & 27.6 \\ 
                    &   & PRF           &  Quad.    & Asymm. & 13.0 \\ \hline
        qa002bo21   & 2.25  & Free      &  --       & Symm. & 96.6    \\
                    &   & BLISS         &  Lin.     & Symm. & -- \\ 
                    &   &  +            &  \textbf{Quad.}    & \textbf{Symm.} & \textbf{0}\\ 
                    &   &  2$^{nd}$     &  --       & Asymm. & 95.3\\ 
                    &   &  order        &  Lin.     & Asymm. & -- \\ 
                    &   &  PRF          &  Quad.    & Asymm. & -- \\ \hline \hline 
        wa052bo11   & 2.25  & Free          &  --       & Symm.  & 1810.2 \\
                    &   & BLISS         &  \textbf{Lin.}     & \textbf{Symm.} & \textbf{0.0}\\ 
                    &   & +             &  Quad.    & Symm.  & -- \\ 
                    &   & 2$^{nd}$      &  --       & Asymm. & 1773.4 \\ 
                    &   & order         &  Lin.     & Asymm. & 134.1 \\ 
                    &   & PRF           &   Quad.   & Asymm. & -- \\ \hline 
        wa052bo12   & 2.25  & Free          &  --       & Symm.  & 533.6 \\
                    &   & BLISS         &  \textbf{Lin.}     & \textbf{Symm.}  & \textbf{0.0} \\ 
                    &   & +             &  Quad.    & Symm.  & -- \\ 
                    &   & 2$^{nd}$      &  --       & Asymm. & 1442.1 \\ 
                    &   & order         &  Lin.     & Asymm. & 740.0 \\ 
                    &   & PRF           &  Quad.    & Asymm. & 674.2 \\ \hline
        wa052bo21   & 2.25  & Free     &  \textbf{--}  & \textbf{Symm.}    & \textbf{0.0}\\
                    &   & BLISS         &  Lin.     & Symm.     & 10.6\\ 
                    &   & +             &  Quad.    & Symm.     & 21.6 \\ 
                    &   & 2$^{nd}$      &  --       & Asymm.    & 22.5 \\ 
                    &   & order         &  Lin.     & Asymm.    & 33.1  \\ 
                    &   & PRF           &   Quad.   & Asymm.    & 44.1 \\ \hline \hline
        wa034bo11   & 2.75  & Free &  \textbf{--}       & \textbf{Symm.} & \textbf{0.0} \\
                    &   & BLISS &  2 Lin.           & Symm. & 24.0 \\ 
                    &   &       &  2 Exp./Lin.      & Symm. & 57.6 \\ 
                    &   &       &  --               & Asymm. & 115.1 \\ 
                    &   &       &  2 Lin.           & Asymm. & 58.3 \\ 
                    &   &       &  2 Exp./Lin       & Asymm. & 80.0 \\ \hline \hline
        wa140bo11   & 2.50  & Fixed &  --       & Symm. & 6.2\\
                    &   & BLISS &  Lin.     & Symm. & 6.5\\ 
                    &   &       &  Quad.    & Symm. & 16.4\\ 
                    &   &       &  \textbf{--}       &\textbf{Asymm.} & \textbf{0.0}\\ 
                    &   &       &  Lin.     & Asymm. & 7.7\\ 
                    &   &       &  Quad.    & Asymm. & 14.9\\ \hline
    \enddata
    \tablecomments{Lin. = Linear Temporal Ramp; Quad. = Quadratic Temporal Ramp.
   Some model combinations result in significantly nonphysical best fits (i.e. negative night side flux or phase offsets $>90^{\circ}$). For further explanation, see text for that planet. Bolded rows denote the best fit.}
\end{deluxetable}
%

\section{Results} \label{results}
In the following sections we detail the individual results for each target, with final best fit values presented in Table \ref{table:Results}.

\subsection{Qatar-1b}
Because the centroids of Qatar-1b fall only partially on the master map (see Figure \ref{fig:centroids}), we reduce the Qatar-1b 4.5 $\micron$ phase curve using both our fixed map method and the standard BLISS map method to compare the resulting phase curves. Recall that partial overlap uses self calibration in non-overlapping regions, but that the overlapping regions provide constraints, particularly when one eclipse overlaps as is the case here.

\begin{figure*}
    \centering
    \includegraphics[width=\textwidth]{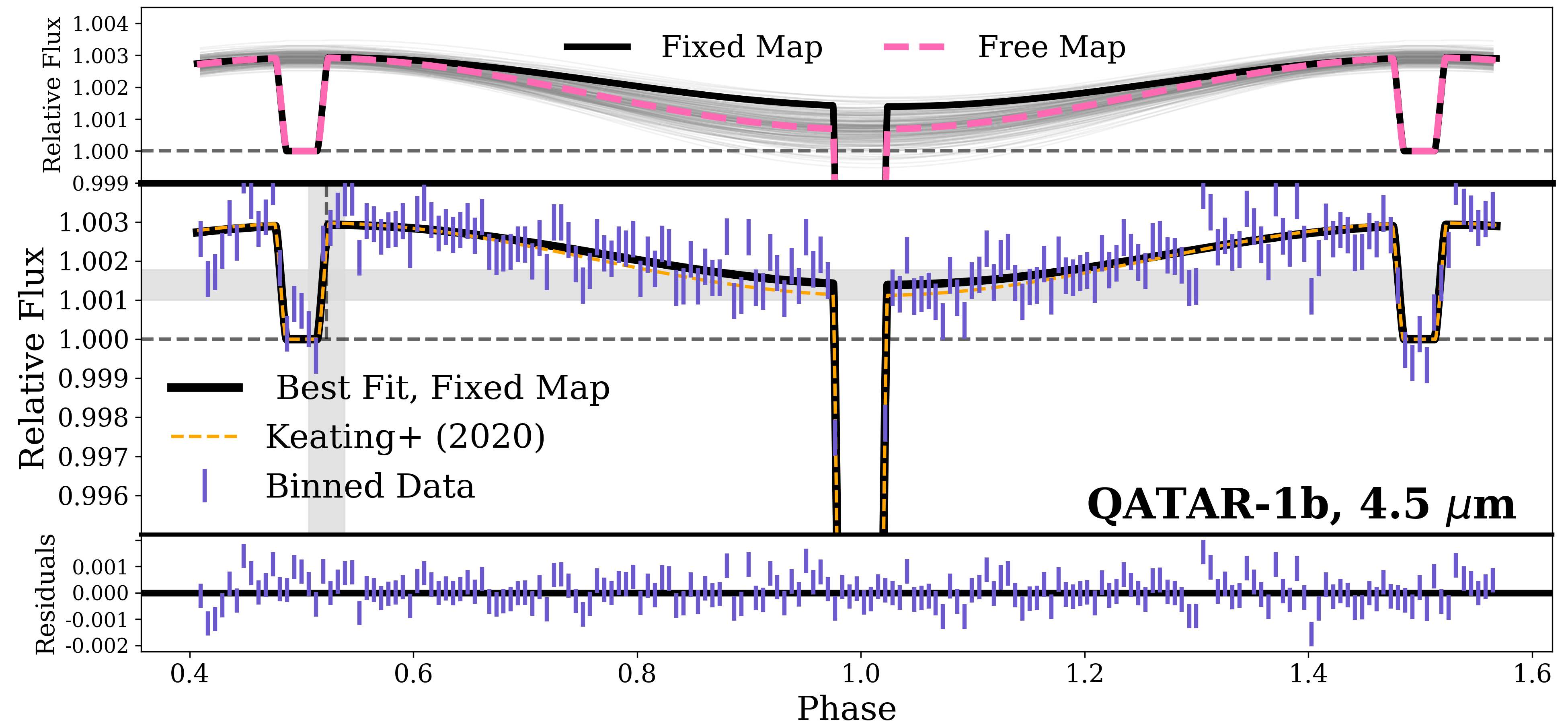}
    \caption{Qatar-1b 4.5 $\micron$ phase curve best fits. \textbf{Top:} a comparison between the fixed and standard (free) BLISS mapping methods, comapred to 200 random draws of the free map fit (note, the fixed map only partially overlaps with the centroids). \textbf{Middle:} Best fit fixed map compared to the binned data. The horizontal shaded regions corresponds to the uncertainty on the minimum flux while the vertical region corresponds to the uncertainty on the phase offset. The residuals are shown in the \textbf{Bottom} panel.  Our results agree well with those from \cite{Keating2020}.}
    \label{fig:Q1_ch2_fit}
\end{figure*}

Figure \ref{fig:Q1_ch2_fit} shows our best fit(s) - the top panel compares our fit with the free map (dashed pink line, standard BLISS method), fixed map (solid black line) compared to 200 random draws of the free map fit to demonstrate that the resulting phase functions are generally consistent. The middle panel shows the best fixed map fit compared to the binned data (data is fit without binning but plotted using bins for ease of presentation). The bottom panel shows the residuals. For this figure, and all following phase curve figures, the vertical shaded region shows the measured offset and associated 1$\sigma$ uncertainties, while the horizontal shaded region denotes the 1$\sigma$ uncertainty on the phase curve minimum.

For Qatar-1b at 4.5 $\micron$, we measure a nightside band integrated brightness temperature of 1098 $\pm$ 158 K and a dayside band integrated temperature of 1696 $\pm$ 39 K. Full results including amplitude, fluxes, offset, eclipse and transit depths are reported in Table \ref{table:Results}.

Both the 4.5 $\micron$ Qatar-1b phase curve, and a 3.6~$\micron$ phase curve that we do not re-fit, were first analyzed by \cite{Keating2020}. The middle panel of Figure \ref{fig:Q1_ch2_fit} compares our best fit to that of \cite{Keating2020}. Both our phase amplitude (calculated based on fluxes) and hot spot offset values agree to within 1-$\sigma$ with a similar slight westward offset as identified by \cite{Keating2020}. 
However, the dayside temperature between our two works is discrepant by 3$\sigma$. This larger difference in temperatures compared to the general agreement between the phase function and eclipse depths is likely a result of different stellar parameters used to estimate the wavelength integrated stellar flux in these channels.

\subsection{Qatar-2b}
For both Qatar-2b data sets, we use the standard BLISS mapping method to remove the intrapixel sensitivity. We also consider 1$^{st}$, 2$^{nd}$, and 3$^{rd}$ order PRF functions for further detrending and find that 2$^{nd}$ order performs best for both channels. 


\subsubsection{3.6 $\micron$} 
In our initial analysis of the 3.6 $\micron$ data set we identify a high frequency sinusoidal systematic. While this may be due to instrumental reasons, we also note that Qatar-2b has been identified as an active star featuring recurring star spots \citep[e.g.][]{Mancini2014,Mocnik2017}, and as such we cannot discount stellar activity. Either way, we chose to model this as an additional source of correlated noise, applying the wavelet methodology of \cite{Carter2009}. This allows us to encapsulate the uncertainty introduced by this high frequency signal into the errors on our reported parameters.  

\begin{figure*}
    \centering
    \includegraphics[width=\textwidth]{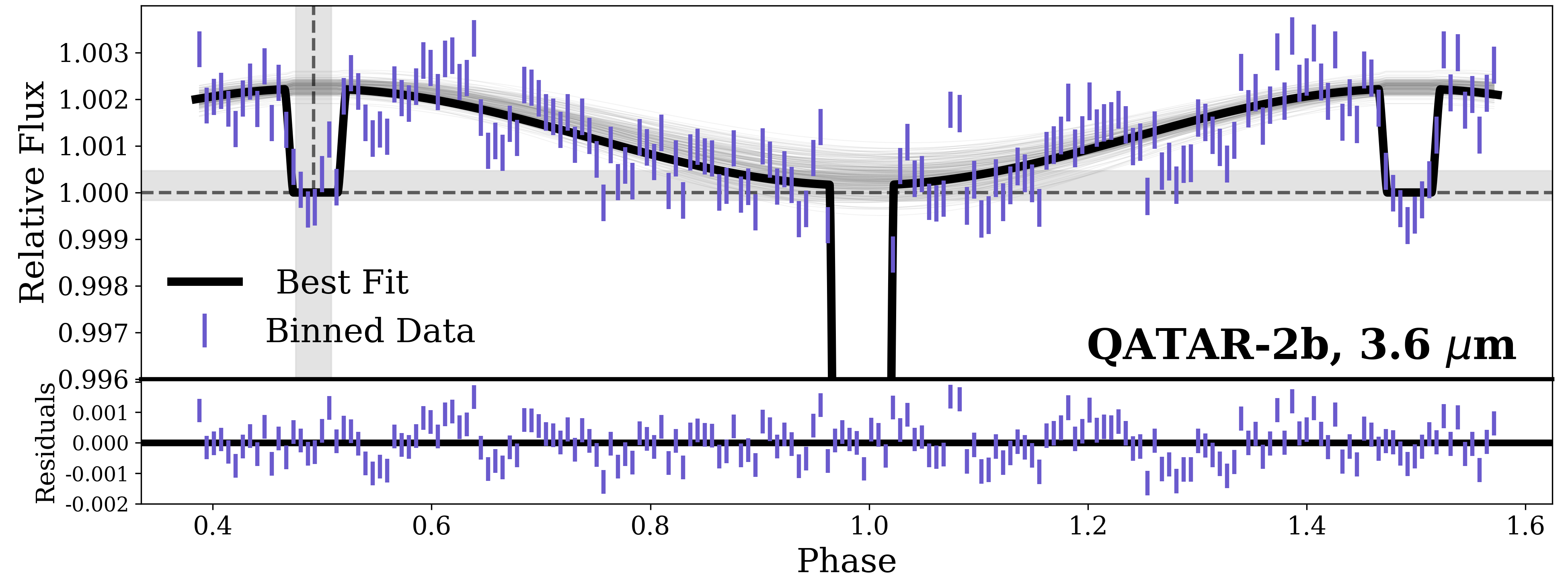}
    \caption{Qatar-2b 3.6 $\micron$ phase curve best fits. \textbf{Top:} Best model combination (Symm. + Quad.) compared to 200 random draws of the best fit, shown as lightly shaded lines. Binned data are overplotted. The horizontal shaded regions corresponds to the uncertainty on the minimum flux while the vertical region corresponds to the uncertainty on the phase offset \textbf{Bottom:} residuals. }
    \label{fig:Q2_ch1_fit}
\end{figure*}

Figure \ref{fig:Q2_ch1_fit} shows our best fit to this data set compared to 200 random draws of the DEMC run. For Qatar-2b at 3.6 $\micron$, we measure a nightside band integrated brightness temperature of 842 $\pm$ 141 K and a dayside band integrated brightness temperature of 1421 $\pm$ 28 K. Full results including amplitude, fluxes, offset, eclipse and transit depths are reported in Table \ref{table:Results}.

\subsubsection{4.5 $\micron$}
The centroids for Qatar-2b do not overlap with our fixed sensitivity map, so we chose to use the standard free BLISS map approach. 

\begin{figure*}
    \centering
    \includegraphics[width=\textwidth]{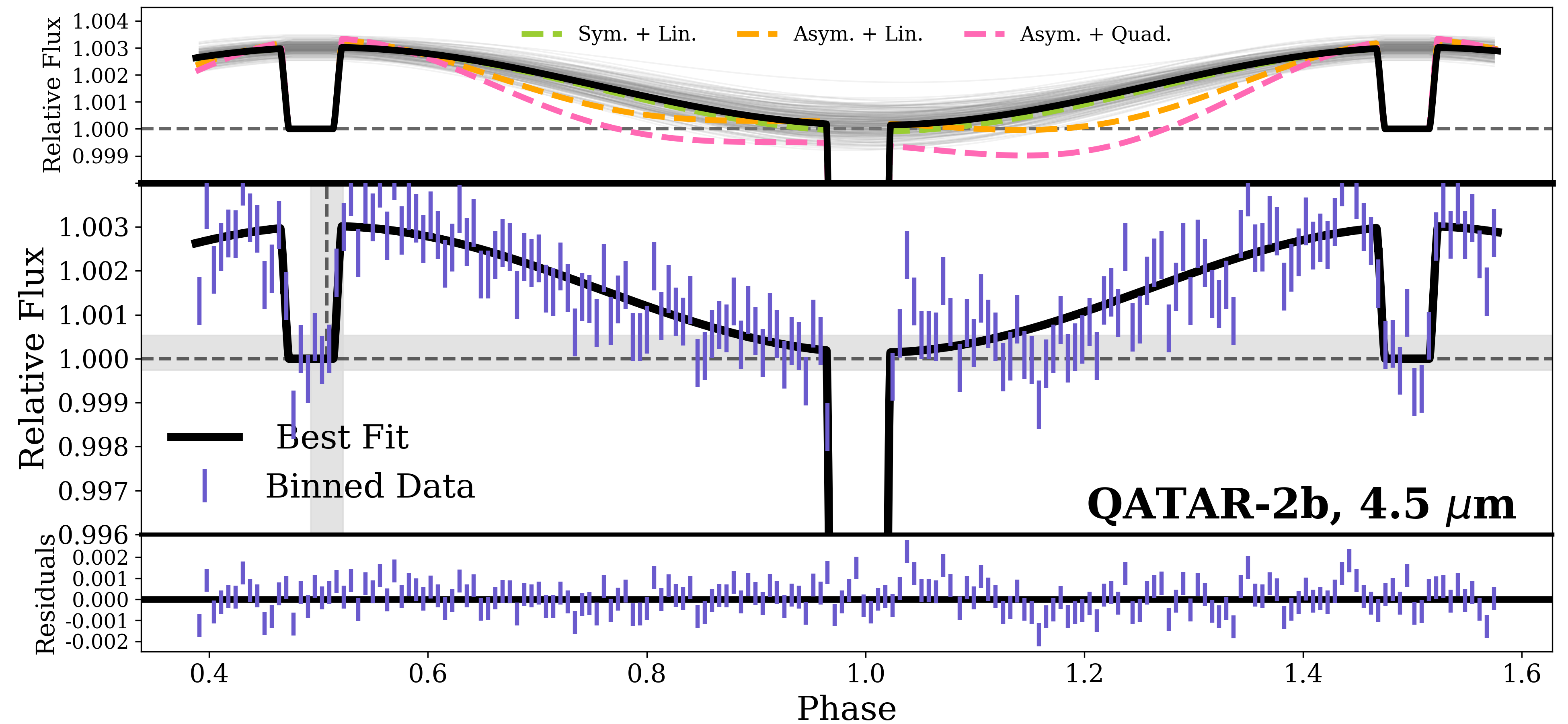}
    \caption{Qatar-2b 4.5 $\micron$ phase curve best fits. \textbf{Top:} Best model combination (Symm. + Quad.) compared to the three other cases which result in a lower BIC but negative nightside fluxes. 200 random draws of the best fit are shown as lightly shaded lines. \textbf{Middle:} Best model compared to the binned data. The horizontal shaded regions corresponds to the uncertainty on the minimum flux while the vertical region corresponds to the uncertainty on the phase offset \textbf{Bottom:} residuals. }
    \label{fig:Q2_ch2_fit}
\end{figure*}

As denoted in Table \ref{table:bestfits}, simple $\Delta$BIC comparisons would suggest that an asymmetric phase function with a quadratic ramp performs best. However, this fit, and the linear ramp option with either a symmetric or asymmetric phase function, result in negative night side flux which is an unphysical result. We therefore consider the symmetric + quadratic ramp as the best fit. We note that all of these model combinations resulted in the same phase offset and the symmetric best-fit amplitudes are within 1$\sigma$. The top panel of Figure \ref{fig:Q2_ch2_fit} shows a comparison of these four model combinations and 200 random draws of our best fit model (solid black line). The general shape of the symmetric + linear combination is relatively similar to that of the quadratic ramp, but the asymmetric functions clearly diverge, which we discuss below. 

Further, as shown in Figure \ref{fig:Q2_ch2_fit}, there is an undetermined systematic after transit that appears in both channels near the jump in centroids at the AOR gap (phase of $\sim$ 1.1-1.2). By eye, it is clear that these asymmetric models are attempting to fit this systematic, providing further weight to our decision to not consider those models. We also find that clipping this region leads to a worse constraint on the BLISS map for this event, with slight modifications in the clipped region heavily impacting the measured phase offset and amplitude. For this reason, and the above discussion, we make the decision not to clip this region. 

As also identified in the 3.6 $\micron$ data for this target, we find a present, but weaker, high frequency sinusoidal systematic. We follow the same approach as at 3.6 $\micron$ to ensure that our errors encapsulate this uncertainty. The middle panel of Figure \ref{fig:Q2_ch2_fit} shows our best fit. 

For Qatar-2b at 4.5 $\micron$ we measure a nightside band integrated brightness temperature of 724 $\pm$ 135 K and a dayside band integrated brightness temperature of 1368 $\pm$ 32 K. Full results including amplitude, fluxes, offset, eclipse and transit depths are reported in Table \ref{table:Results}.

\begin{figure*}
    \centering
    \includegraphics[width=\textwidth]{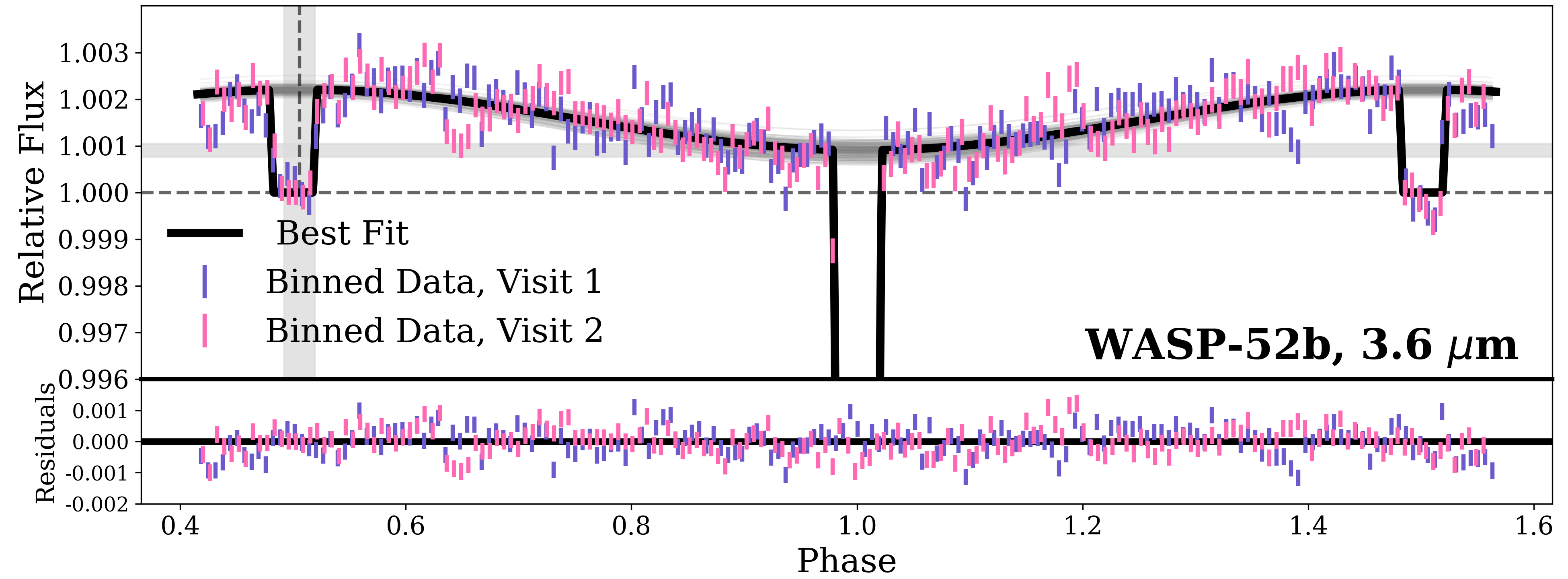}
    \caption{WASP-52b 3.6 $\micron$ phase curve best fit. \textbf{Top:} Best model combination (Symm. + Lin.) compared to 200 random draws of the best fit, shown as lightly shaded lines. Binned data are overplotted. The horizontal shaded regions corresponds to the uncertainty on the minimum flux while the vertical region corresponds to the uncertainty on the phase offset. \textbf{Bottom:} residuals. }
    \label{fig:W52_ch1_fit}
\end{figure*}

\begin{figure*}
    \centering
    \includegraphics[width=\textwidth]{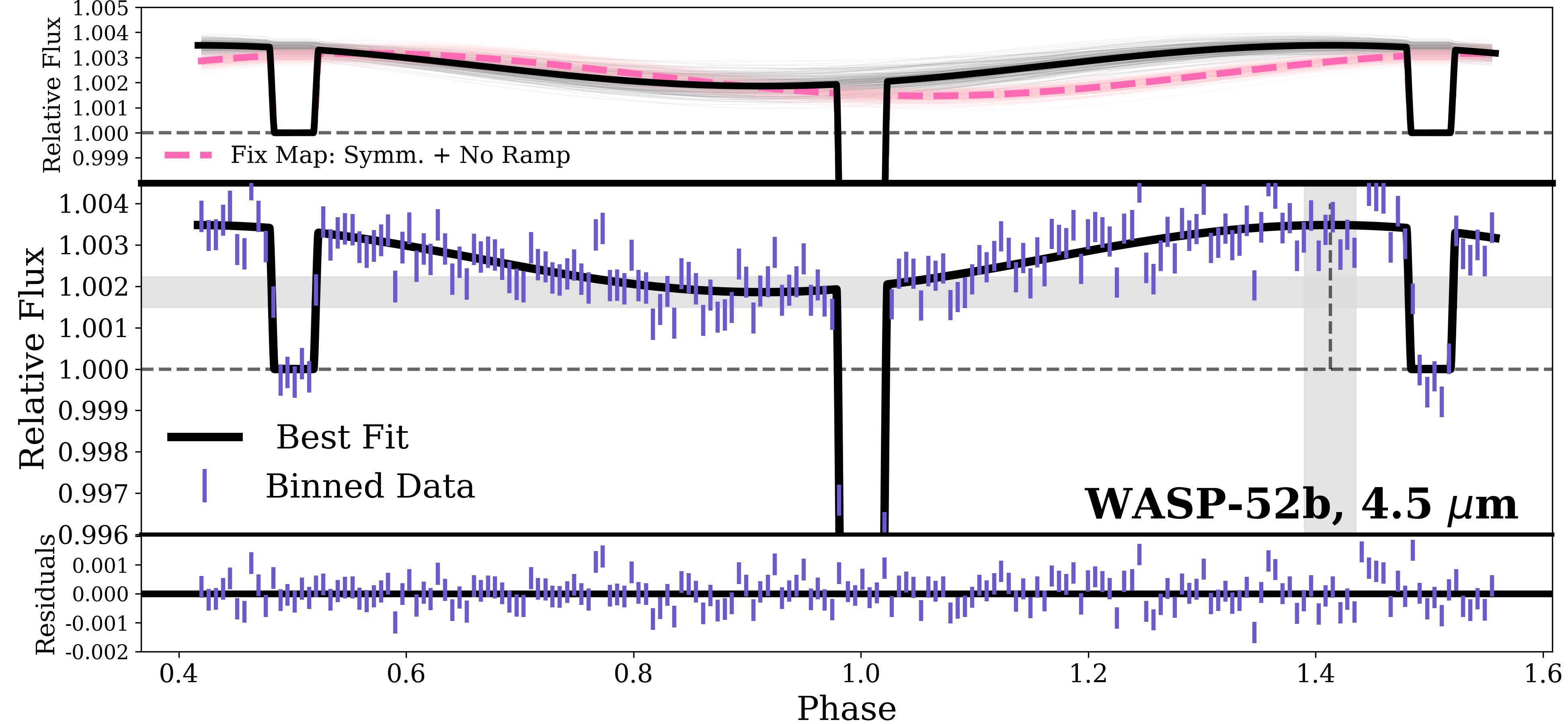}
    \caption{WASP-52b 4.5 $\micron$ phase curve best fit. \textbf{Top:} Best model combination (Symm. + No Ramp) compared to 200 random draws of the best fit shown as lightly shaded lines in black. The fixed map best fit is shown as a pink dashed line. \textbf{Middle:} Best model compared to the binned data. The horizontal shaded regions corresponds to the uncertainty on the minimum flux while the vertical region corresponds to the uncertainty on the phase offset. \textbf{Bottom:} residuals. }
    \label{fig:W52_ch2_fit}
\end{figure*}

\subsection{WASP-52b}
\subsubsection{3.6 $\micron$}
Two phase curves of WASP-52b were observed at 3.6 $\micron$. As shown in Figure \ref{fig:BLISSmaps}, both observations have two non-overlapping groups of centroids, which makes removing the intrapixel effect particularly difficult. We found that individually, the phase curve parameters were poorly constrained due to a degeneracy between pixel position and phase function parameters. To mitigate this as best as possible, we performed a joint fit of both visits with all astrophysical signal parameters tied to each other between visits, while the systematics are individually fit to account for time variability in the 3.6 $\micron$ intrapixel response \citep[3.6 $\micron$ sensitivity variability is discussed in][]{May2020}. In Table \ref{table:bestfits} we include the $\Delta$BICs for both events from a given joint fit. Because of the unphysical negative nightside flux resulting from the use of a quadratic ramp, we select the linear temporal ramp with a symmetric phase function as our best fit. Figure \ref{fig:W52_ch1_fit} shows our best fit phase function compared to both data sets as well as 200 random draws from our MCMC chains.

For WASP-52b at 3.6 $\micron$ we measure a nightside band integrated brightness temperature of 1116 $\pm$ 46 K and a dayside band integrated brightness temperature of 1454 $\pm$ 21 K. Full results including amplitude, fluxes, offset, eclipse and transit depths are reported in Table \ref{table:Results}.

\subsubsection{4.5 $\micron$}
The 4.5 $\micron$ phase curve of WASP-52b partially overlaps with our fixed sensitivity map. For this data set, the long swipe of centroids centered at a y-position of $\sim$14.2 represent centroids that are never revisited, leaving a strong degeneracy between astrophysical and systematic signals (when centroids slowly drift and do not remain fairly constant, the change in flux due to changing centroids can mimic visit long trends like the planet's phase curve). Because of this and a larger-than-normal effect of the PRF Gaussian widths, we find that this data set is not well modeled by the fixed sensitivity map. As a result, we use a standard BLISS map approach and find that a 2$^{nd}$ order PRF-FWHM produces the best results. Our best fit adopts no temporal ramp with a symmetric phase function. 

The top panel of Figure \ref{fig:W52_ch2_fit} shows our best fit free map (solid black line) compared to our best fit fixed map (dashed pink) and 200 draws of the best free map fit. When using the fixed map, we find that this degeneracy between the intrapixel sensitivity and phase function results in the phase curve offset being strongly dependent on the choice of temporal ramp. Notably, a linear ramp results in inverted phase curve (i.e. a phase offset of 180$^{\circ}$, suggesting the nightside is the hottest, an unphysical result) while a quadratic ramp results in a phase offset of 75$^{\circ}$, also a result that is unexpected from 3D models which predict significantly smaller offsets. All ramps with an asymmetric phase function place the hot spot squarely on the nightside of the planet with offsets $>100^{\circ}$. With these offsets in strong disagreement with the measured 3.6 $\micron$ offset, we conclude that the fixed map is not appropriate for such strongly drifting centroids due to secondary systematics this introduces. The middle panel of Figure \ref{fig:W52_ch2_fit} shows our best fit compared to the data, with residuals in the bottom panel.

For WASP-52b at 4.5 $\micron$ we measure a nightside band integrated brightness temperature of 1224 $\pm$ 77 K and a dayside band integrated brightness temperature of 1481 $\pm$ 34 K. Full results including amplitude, fluxes, offset, eclipse and transit depths are reported in Table \ref{table:Results}. We caution that this data set is potentially unreliable and is worthy of follow up with future missions.

\begin{figure*}
    \centering
    \includegraphics[width=\textwidth]{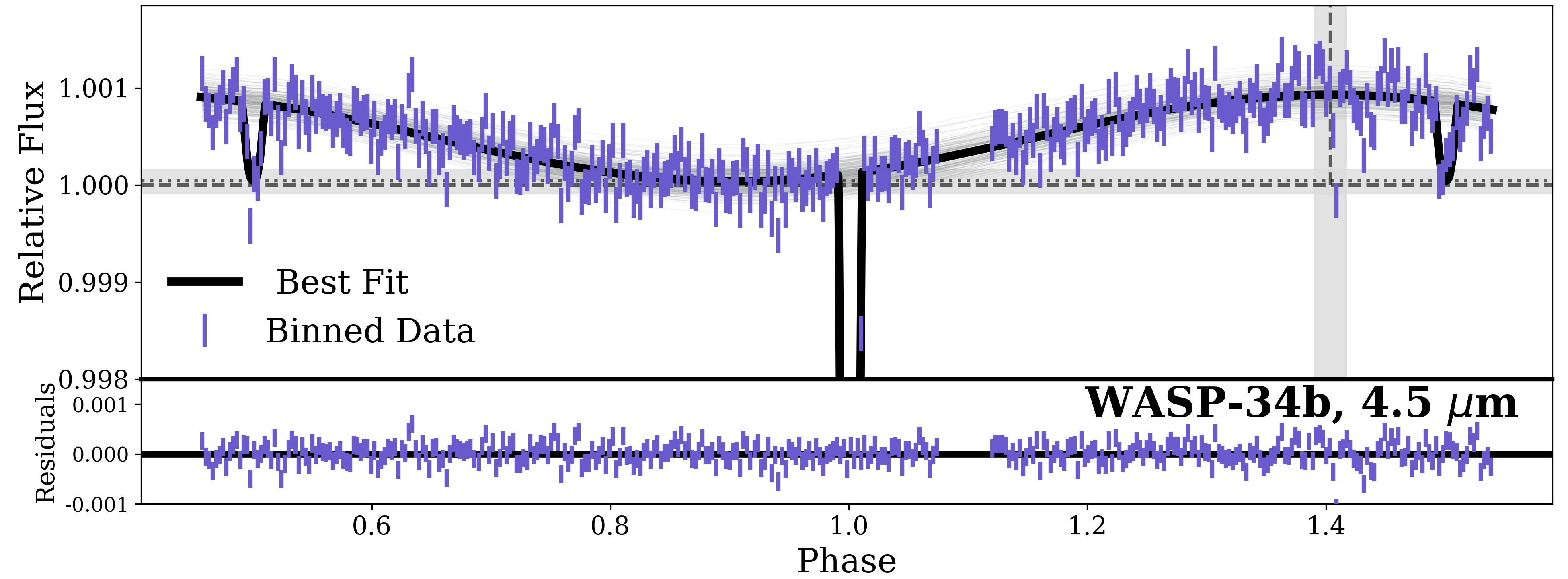}
    \caption{WASP-34b 4.5 $\micron$ phase curve best fit. \textbf{Top:} Best model combination (Symm. + No Ramp) compared to 200 random draws of the best fit, shown as lightly shaded lines. Binned data are overplotted. The horizontal shaded regions corresponds to the uncertainty on the minimum flux while the vertical region corresponds to the uncertainty on the phase offset. \textbf{Bottom:} residuals. The dashed horizontal line denotes the stellar flux level, the dotted horizontal line denotes the in-eclipse flux which is higher due to the grazing event. The gap around a phase of 1.1 is from a data downlink gap due to the 4.3 day orbital period.}
    \label{fig:W34_ch2_fit}
\end{figure*}

\begin{figure*}
    \centering
    \includegraphics[width=\textwidth]{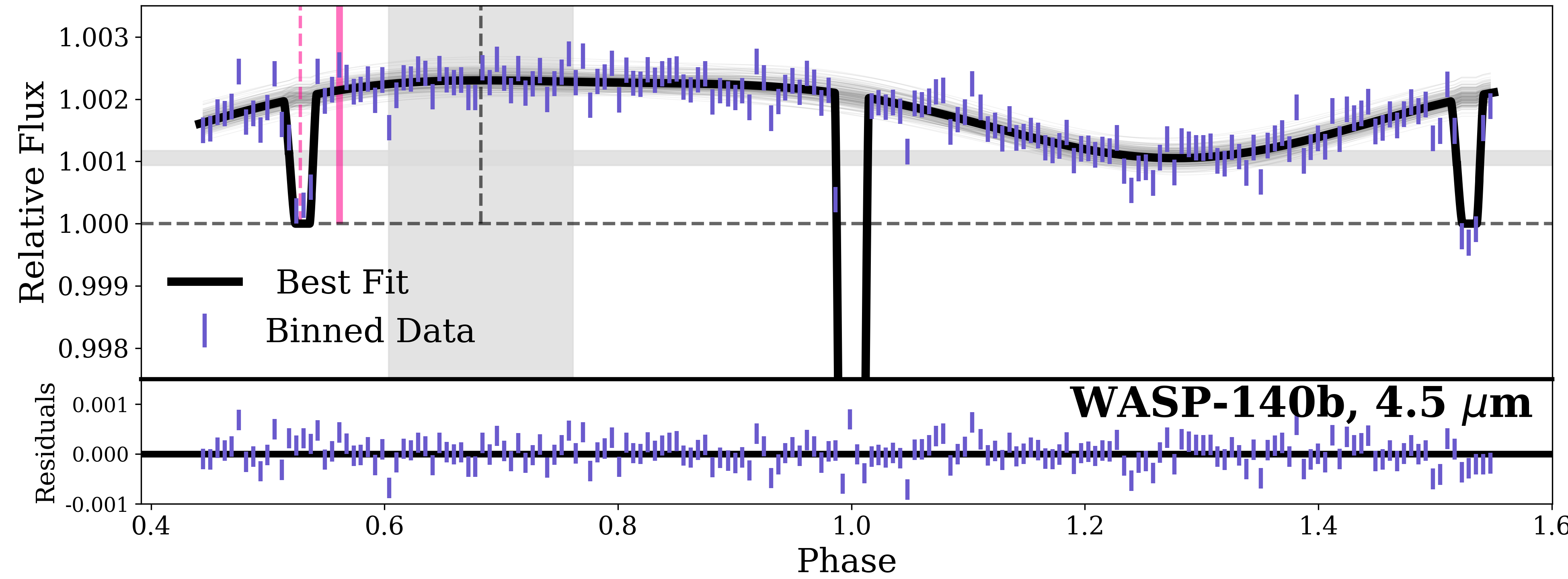}
    \caption{WASP-140b 4.5 $\micron$ phase curve best fit. \textbf{Top:} Best model combination (Symm. + No Ramp) compared to 200 random draws of the best fit, shown as lightly shaded lines. Binned data are overplotted. The horizontal shaded regions corresponds to the uncertainty on the minimum flux while the vertical region corresponds to the uncertainty on the phase offset. \textbf{Bottom:} Residuals. As a reminder, WASP-140b has a slight eccentricity, possibly explaining the large westward offset. The solid pink line denotes periastron as compared to the secondary eclipse (dashed pink line).}
    \label{fig:W140_ch2_fit}
\end{figure*}

\subsection{WASP-34b}
The 4.5 $\micron$ phase curve of WASP-34b is one of the longest phase curve observed  by \spitzer\, at 112.4 hours total over 5 AORs. As a result, there is a brief down-link gap after transit and an associated exponential ramp and large drifts in centroid position after WASP-34 was reacquired. Because the telescope had to move during the course of the phase curve, we fit the data before and after the gap with independent temporal ramps. In addition, the large drifts in centroid position associated with the start of the first AOR and the first AOR after the gap, result in strong exponential trends in flux. These exponential changes in measured flux are poorly constrained by our fixed sensitivity map suggesting the instrument needed longer stabilization times both at the start of the observations and after the downlink gap. Therefore, because of the unique situation this data gap causes, we do not apply the fixed sensitivity map, and instead use standard BLISS mapping to self-calibrate the data. 

For the two component temporal ramp we test (1) an exponential + linear ramp combination to address the initial large drifts, (2) a linear ramp, and (3) no temporal ramp. For each case, the ramp is split into two components; before and after the down-link gap. The data is not well modeled with quadratic ramps. As highlighted in Table \ref{table:bestfits}, we select `no ramp' as the best ramp model in combination with a symmetric phase function. Figure \ref{fig:W34_ch2_fit} shows our best fit phase curve in combination with the binned data, residuals are shown in the bottom panel. Recall that the vertical shaded region shows the peak of the phase curve and associated uncertainty, while the horizontal shaded region corresponds to the uncertainty on the minimum of the phase curve.

To correct for the grazing event, we calculated the percentage of planet that overlaps with the star assuming an impact parameter of 0.904 from \cite{Smalley2011}. We then renormalize the phase curve based on our calculated 47 $\pm$ 6 ppm contamination from the dayside of WASP-34b during eclipse. All reported values have been corrected for this. 

Our best fit results for WASP-34b at 4.5 $\micron$ give a nightside band integrated brightness temperature of 726$\pm$119 K and a dayside band integrated brightness temperature of 1184 $\pm$ 47 K. Full results including amplitude, fluxes, offset, eclipse and transit depths are reported in Table \ref{table:Results}. All values have been corrected to account for the grazing transit



\subsection{WASP-140b}
The 4.5 $\micron$ phase curve of WASP-140b was split into three AORs, each resulting in individual independent regions of centroids, as shown in Figure \ref{fig:centroids} and Figure \ref{fig:BLISSmaps}. As has been seen in other \spitzer\ data sets, independent groups of centroids result in stronger degeneracies between the systematic and astrophysical models. Luckily, each of the three groups of centroids at least partially overlaps with our fixed sensitivity map, thus improving our ability to remove this degeneracy. 

For comparison, we also fit the WASP-140b data set without our fixed sensitivity map (i.e. using a standard BLISS approach). We found the results with a free map preferred flat (amplitude of zero) phase curves with longer than normal MCMC convergence times, suggesting the phase space of `allowed' fits is large. While typically our results both with and without our fixed sensitivity map are similar in shape or general phase curve parameters, WASP-140b is a case in which an underlying phase curve could only be determined with the use of our fixed map. This is likely due to the three independent regions of centroids which results in the phase function easily being modeled away by systematics.

Figure \ref{fig:W140_ch2_fit} shows our best fit compared to 200 random draws from the MCMC chains. Our best fit model for WASP-140b includes no temporal ramp and an asymmetric phase function with a large westward phase offset of -55.7$^{\circ}$ $\pm$ 28$^{\circ}$. 

Eastward offsets are expected for synchronously rotating hot Jupiters. However, a westward phase offset of -23 $\pm$ 4$^{\circ}$ measured for CoRoT-2b \citep{Dang2018} suggests that a large westward offset is not a newly observed or unique phenomenon. One possible explanation for WASP-140b's measured westward offset is that it is non-synchronously rotating, which is supported by its eccentric orbit (e = 0.047 $\pm$ 0.0035). \cite{Hellier2017} suggest that WASP-140b may have only recently arrived at its current, short-orbital-period location due to the system's short circularization time. \added{3D circulation models predict that such rotation rates (i.e. slower than synchronous) can shift the hotspot westward \citep[e.g.][]{Rauscher2014}. While magnetic effects and deep jets can also produce westward-to-no phase offsets \citep[e.g.][]{Rogers2014,Carone2020}, the equilibrium temperature of WASP-140b below the threshold where magnetic effects are important.} An additional explanation is that the peak of the phase curve simply corresponds to periastron passage, which occurs after secondary eclipse for WASP-140b, as denoted in Figure \ref{fig:W140_ch2_fit}, affecting the shape and peak location of the observed phase curve \citep[][,albeit for significantly higher eccentricities]{Lewis2010,Lewis2013,Kataria2013,Mayorga2021}. Phase curves of the eccentric planets XO-3b \citep[e = 0.29][]{Dang2022}, WASP-14b \citep[e=0.08][]{Wong2015}, and HAT-P-2b \citep[e = 0.51]{Lewis2013} have also been observed by \spitzer. We leave a further discussion of the shape of the WASP-140b phase curve and it's comparisson to other eccentric planets for future work.

Our best fit results for WASP-140b at 4.5 $\micron$ give a nightside band integrated brightness temperature of 1201 $\pm$ 29 K and a dayside band integrated brightness temperature of 1169 $\pm$ 25 K. Full results including amplitude, fluxes, offset, eclipse and transit depths are reported in Table \ref{table:Results}. We note that the nightside (phase 0.0/1.0) appears warmer than the dayside (phase 0.5) because of the large phase offset.
%

\begin{sidewaystable}[]
    \tabletypesize{\footnotesize}
    \caption{Fit Results}
    \label{table:Results}
    \renewcommand\arraystretch{1.1}
    \tablecolumns{10}
    \begin{center}
    \begin{tabular}{|r l|c|c|c|c|c|c|c|c|}
    \hline
        \multicolumn{2}{|r|}{ } & qa001bo21 & qa002bo11 & qa002bo21 & \multicolumn{2}{|c|}{wa052bo11} & wa052bo21 & wa034bo21* & wa140bo21 \\
          \multicolumn{2}{|r|}{ } & { } & {} & {} & \multicolumn{2}{|c|}{wa052bo12} & {} & {}& {}  \\ \hline \hline
        Eclipse Depth & [ppm]           & 2914$\pm$162          & 2223$\pm$118 & 3004$\pm$ 185 & \multicolumn{2}{|c|}{2201$\pm$77} & 3350$\pm$162 & 850$\pm$95 & 2022$\pm$111 \\ \hline
        Transit Depth & [R$_p$/R$_S$]   & 0.14514$\pm$0.00018   & 0.16369$\pm$0.00011 & 0.16189$\pm$0.00015 & \multicolumn{2}{|c|}{0.16464$\pm$0.00011} & 0.16305$\pm$0.00014 & 0.12007$\pm$0.0002& 0.17399$\pm$0.00012 \\ \hline
        Hotspot Offset & [$^{\circ}$]   & -7.98$\pm$5.79 & 0.6$\pm$5.8 & -5.2$\pm$5.4 & \multicolumn{2}{|c|}{-2.1$\pm$4.9} & 31.8$\pm$7.8 & 34.7$\pm$4.7 & -55.7$\pm$28.3 \\ \hline
        Amplitude & [ppm]  & 769$\pm$213           & 1039$\pm$165 & 1442$\pm$221 & \multicolumn{2}{|c|}{650$\pm$79} & 812$\pm$204 & 446$\pm$78 &  624$\pm$80\\ \hline
        \multirow{4}{*}{F$_p$/F$_S$ [ppm]}  & \multicolumn{1}{|l|}{Max} & 2930$\pm$164  & 2228$\pm$119 & 3019$\pm$193 & \multicolumn{2}{|c|}{2207$\pm$77} & 3483$\pm$181 & 930$\pm$97 & 2302$\pm$111 \\ 
                    \cline{2-10}            & \multicolumn{1}{|l|}{Min} & 1392$\pm$394  & 151$\pm$ 308 & 136$\pm$399 & \multicolumn{2}{|c|}{906$\pm$138} & 1860$\pm$366 & 38$\pm$122& 1053$\pm$114 \\ 
                    \cline{2-10}            & \multicolumn{1}{|l|}{Day} & 2922$\pm$163  & 2227$\pm$ 118 & 3018$\pm$186 & \multicolumn{2}{|c|}{2206$\pm$77} & 3365$\pm$163 & 851$\pm$95 & 1904$\pm$111\\ 
                    \cline{2-10}            & \multicolumn{1}{|l|}{Night} & 1399$\pm$213 & 152$\pm$ 309 & 138$\pm$400 & \multicolumn{2}{|c|}{906$\pm$138} & 1978$\pm$357 & 117$\pm$123 & 2064$\pm$132 \\ \hline
        \multirow{4}{*}{Temperature [K]}  & \multicolumn{1}{|l|}{Max}   & 1697$\pm$39 & 1422$\pm$ 28 & 1370$\pm$33 & \multicolumn{2}{|c|}{1455$\pm$21} & 1504$\pm$36 & 1222$\pm$47 & 1252$\pm$25 \\ 
                    \cline{2-10}            & \multicolumn{1}{|l|}{Min} & 1096$\pm$157 & 840$\pm$141 & 720$\pm$134 & \multicolumn{2}{|c|}{1115$\pm$45} & 1195$\pm$81 & 666$\pm$117 & 973$\pm$32 \\ 
                    \cline{2-10}            & \multicolumn{1}{|l|}{Day} & 1696$\pm$39 & 1421$\pm$28 & 1368$\pm$32 & \multicolumn{2}{|c|}{1454$\pm$21} & 1481$\pm$34 & 1185$\pm$47 & 1169$\pm$25 \\ 
                    \cline{2-10}            & \multicolumn{1}{|l|}{Night} & 1098$\pm$158 & 842$\pm$141 & 724$\pm$135 & \multicolumn{2}{|c|}{1116$\pm$46} & 1224$\pm$77 & 726$\pm$119 & 1201$\pm$29 \\ \hline
         
    \end{tabular}
    \end{center}
    \tablecomments{Label denotes the planet (e.g. qa001b = Qatar-1b), type of observation (o=orbit), \spitzer\ IRAC channel (1 or 2) and visit number (1 or 2). "Max" and "Min" refer to the maximum and minimum of the phase curve. "Day" and "Night" refer to a phase of 0.0/1.0 and 0.5, respectively. A positive phase offset corresponds to an eastward shift. Amplitude is given as (max-min)/2.}
\end{sidewaystable}

\begin{figure}
    \centering
    \includegraphics[width=0.5\textwidth]{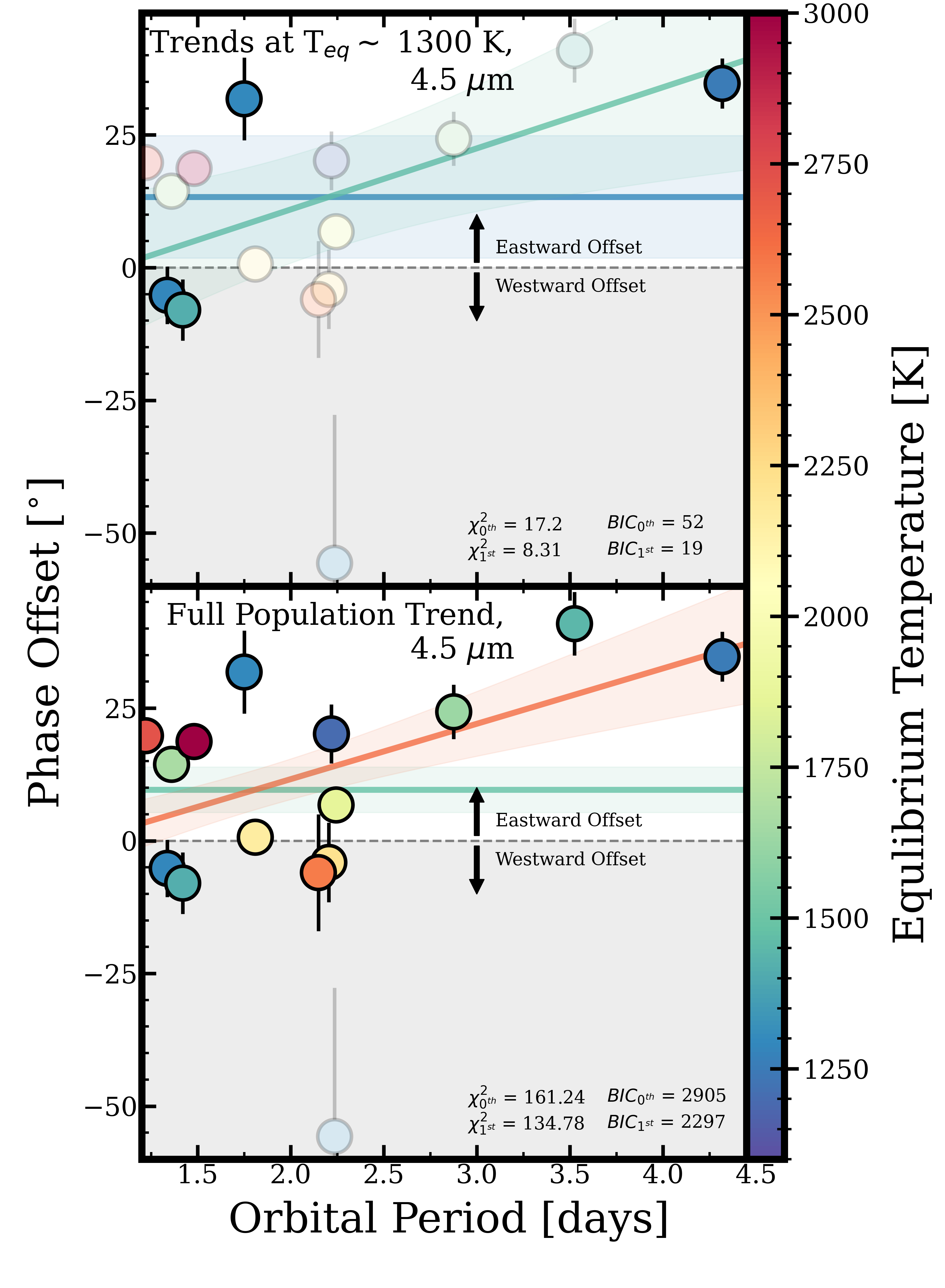}
    \caption{Phase offset at 4.5 $\micron$ vs. orbital period. \textbf{Top:} trend lines for T$_{eq}\sim$ 1300 K sample only. Points outlined in black are the planets analyzed in this work (i.e. the 1300 K sample). The remaining sample is shown in the background. \textbf{Bottom:} trend lines for the complete sample of published hot Jupiter phase curves. For both panels, the color of each point corresponds to the equilibrium temperature. Over plotted are a first order and zeroth order trend lines. In both cases we find the first order linear trend best explains the observations.}
    \label{fig:phase_v_period}
\end{figure}

\section{Population Trends} \label{comp}
Extracting robust population level trends in hot Jupiters from \spitzer\ phase curve observations has proved difficult due to inconsistencies in data reduction and lack of consistent reporting of phase curve and stellar parameters used to derive day side and night side temperatures, for example. This is commonly seen in large scatter in observed trends, as well as the disagreement between model trends and observed trends (see discussion in introduction). In this work we present a uniform sample for seven of the planets (the five from this work, as well as WASP-76b from \citetalias{MayKomacek2021}, \citeyear{MayKomacek2021} and WASP-43b from \citealt{May2020}) observed and analyzed as uniformly as possible, including adopting equilibrium temperatures calculated with the same stellar parameters used to convert planet-star flux ratios to brightness temperatures. This approach reduces potential bias as much as possible. \added{The remainder of the population includes results from, in no particular order, HD 189733b \citep{Knutson2012}, HD 209458b \citep{Zellem2014}, HAT-P-7b and WASP-19b \citep{Wong2016}, HD 149026b and WASP-33b \citep{Zhang2018}, WASP-18b \citep{Maxted2013}, WASP-12b \citep{Bell2019}, WASP-14b \citep{Wong2015}, WASP-103b \citep{Kreidberg2018}, KELT-16b and MASCARA-1b \citep{Bell2021}, and KELT-9b \citep{Mansfield2020}.} For these remaining planets in the \spitzer\ sample, it is not always clear what stellar and planetary parameters were used, so we adopt the equilibrium temperature reported in the composite parameters table of the NASA Exoplanet archive, \added{exact parameters used can be found on our continuously updated repository of results (see link at the end of Section \ref{intro}.} Future work reanalyzing the literature sample will use consistent values.

Here we present evidence for a trend of increasing phase offset with orbital period at 4.5 $\micron$, as shown in \cite{Parmentier2018}. In Figure \ref{fig:phase_v_period} we show measured 4.5 $\micron$ hotspot offsets vs. orbital period. The points outlined in black in the top panel are four of the five planets from this work, all with $T_{eq}\sim$ 1300 K. WASP-140b is excluded from trend fitting due to its eccentric orbit and resulting anomalous offset. We find statistical evidence for an increasing trend of offset with period ($\Delta$BIC = -33 as compared to just a flat line) for the remaining four planets in our T$_{eq}\sim$1300 K sample (top), and when considering the entire set of literature values in addition to our new results (bottom), we find even stronger evidence for this linear trend of increasing offset with orbital period ($\Delta$BIC = -608 as compared to just a flat line). We note, however, that this trend relies heavily on the three observed planets with orbital periods greater than 2.5 days. 

The observational evidence for the existence of the offset vs. orbital period trend is dynamically interesting because theory does not predict a dependence on rotation rate (which is related to orbital period for tidally locked planets). While \cite{Hammond2018} show that the offset does depend on a non-dimensional parameter $G$, which is a function of scale height, gravity, radius, and rotation rate, it is not clear that rotation rate is the sole driver of that effect. In fact,   \cite{Zhang2017} predict that the hot spot offset is solely driven by the ratio between the radiative and advective timescale. Because rotation rate should not affect the radiative timescale, the existence of this trend suggest that rotation rate has implications on the wind speed in tidally locked hot Jupiter atmospheres, namely that shorter period (i.e. faster rotation rates) planets must have weaker equatorial jets. This trend can therefore be confirmed with ground-based high resolution spectroscopy that directly probes wind speeds in hot Jupiter atmospheres. 

We also explore the potential for this trend of increasing phase offset with orbital period at 3.6 $\micron$. The lack of reliable offset measurements past orbital periods of $\sim$2.25 days at 3.6 $\micron$ precludes our ability to definitely detect such a trend specific to this channel; however, the measured offsets at 3.6 $\micron$ are statistically consistent with the 4.5 $\micron$ trend.

\begin{figure}
    \centering
    \includegraphics[width=0.5\textwidth]{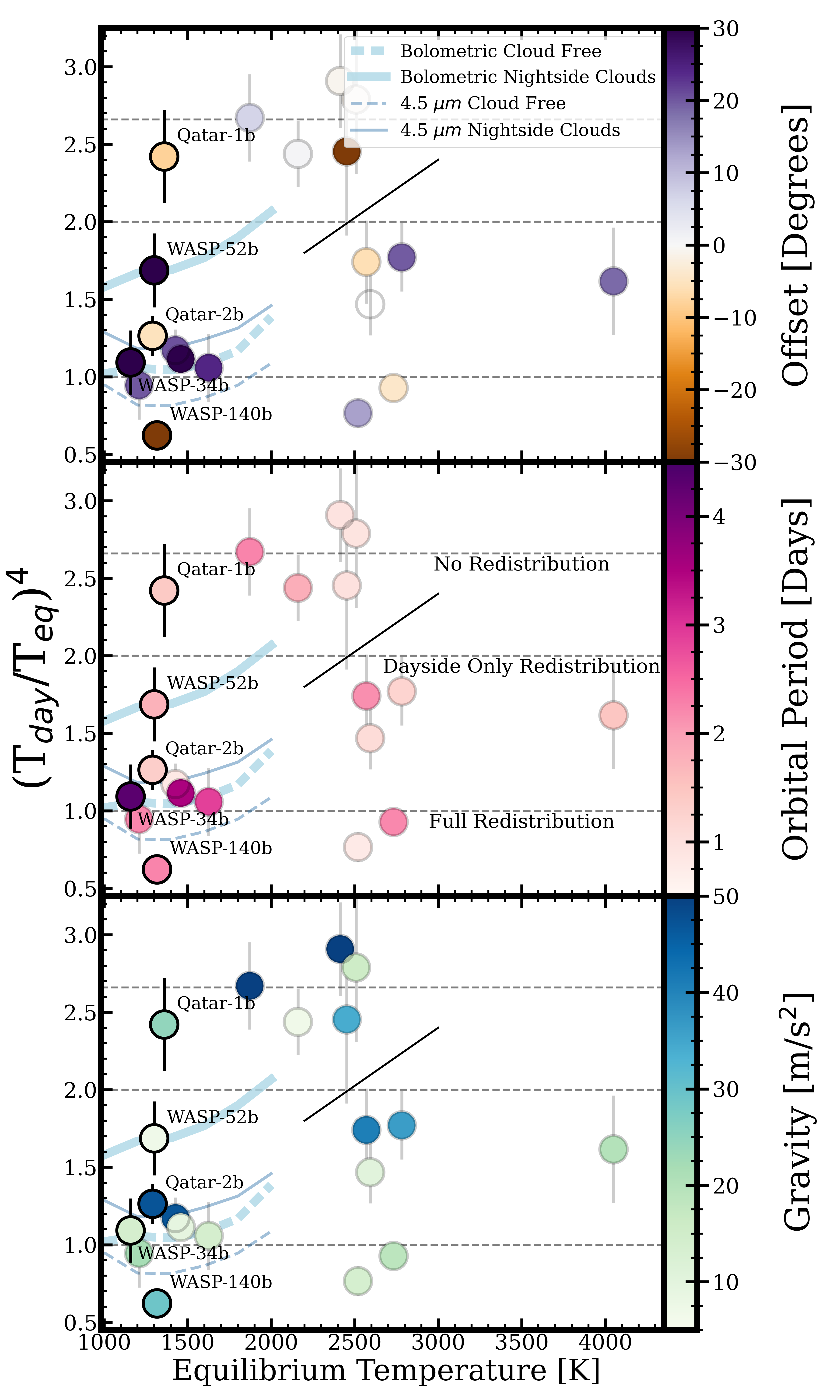}
    \caption{Apparent heat redistribution vs. equilibrium temperature compared to three different tertiary parameters: phase offset, orbital period, and gravity (shown with the color scales) for the 4.5 $\micron$ data. Models are from \cite{Parmentier2021} showing both 4.5 $\micron$ and bolometric emission (thin and thick lines, respectively) for both cloudy and clear (sold and dashed lines, respectively) cases. We note that some observations fall above even the bolometric predictions, suggesting a departure from current theory. Those points with dark outlines and error bars are our $\sim$ 1300 K sample. The remaining points are from literature values. From top to bottom, horizontal dashed lines denote values expected from no heat redistribution, dayside only redistribution, and full redistribution. A solid line denotes an approximate division between literature planets with little heat distribution and those with a significant amount. There is no clear trend with offset, orbital period, or gravity.}
    \label{fig:redistribution}
\end{figure}

Additionally, we explore the relationship between the apparent heat redistribution (T$_{b}$/T$_{eq}$)$^4$ versus the equilibrium temperature, as shown in Figure \ref{fig:redistribution}. This relationship between dayside brightness temperatures and the equilibrium temperature is dependent on the circulation efficiency and the planet's albedo, and has been predicted by \cite{Cowan2011b, Schwartz2017}, and \cite{Parmentier2021}. In Figure \ref{fig:redistribution}, we show the apparent heat redistribution parameter at 4.5 $\micron$ vs. the equilibrium temperature, compared to three secondary parameters denoted with color in the three panels. For the planets in our sample (darkly outlined points) we use the same stellar parameters to calculate the dayside brightness temperature and the equilibrium temperature for consistency (see Tables \ref{table:planet params} and \ref{table:LimbDarkening}). However, because we do not know the stellar parameters used for most of the literature sample, there may be biases in the plotted literature (lightly outlined points) sample. We note that biases due to the stellar parameters are unlikely to explain all the scatter here due that seen in our uniform sample alone (see below).

Figure \ref{fig:redistribution} shows lines corresponding to expected (T$_{day}$/T$_{eq}$)$^{4}$ ratios for no heat redistribution (2.67), dayside only redistribution (2.0), and full heat redistribution (1.0) from models, as well as trend lines from \cite{Parmentier2021} for bolometric and 4.5 $\micron$ expectations both with and without nightside clouds for T$_{eq} <$ 2000 K. Because other 3D models predict nightside clouds primarily dissipate above $\sim$ 2000K \citep[e.g.][]{Roman2021}, we do not expect that a simple extrapolation of these trend lines to higher temperatures is appropriate, particularly because we would expect the cloudy and cloud-free models to converge \citep{Roman2021}. However, even with those caveats, we still see that approximately one-third of the literature sample fall above even the bolometric emission models from \cite{Parmentier2021}, suggesting even warmer daysides than expected from current theory. In our $\sim$1300 K sample, the hotter than expected dayside temperatures are further interesting due to their equilibrium temperature being below the threshold where MHD effects begin to affect the day-night heat transport \citep{Menou2012, Rogers2014, Hindle2021a, Hindle2021b}. As a result, unconsidered MHD effects cannot explain why Qatar-1b and WASP-52b have warmer than expected day sides compared to the \citeauthor{Parmentier2021} models.

Importantly, we see a strong spread within our $\sim$1300 K sample (mean of (T$_{\rm{b}}$/T$_{\rm{eq}}$)$^4$ = 1.42 with a standard deviation of 0.61), suggesting that secondary parameters are as important in predicting the heat transport efficiency in hot Jupiters as equilibrium temperature. We do not see any secondary trends in offset, orbital period (as a proxy for rotation rate, assuming tidally locked planets), or gravity, as shown in Figure \ref{fig:redistribution}.

\begin{figure}
    \centering
    \includegraphics[width=0.5\textwidth]{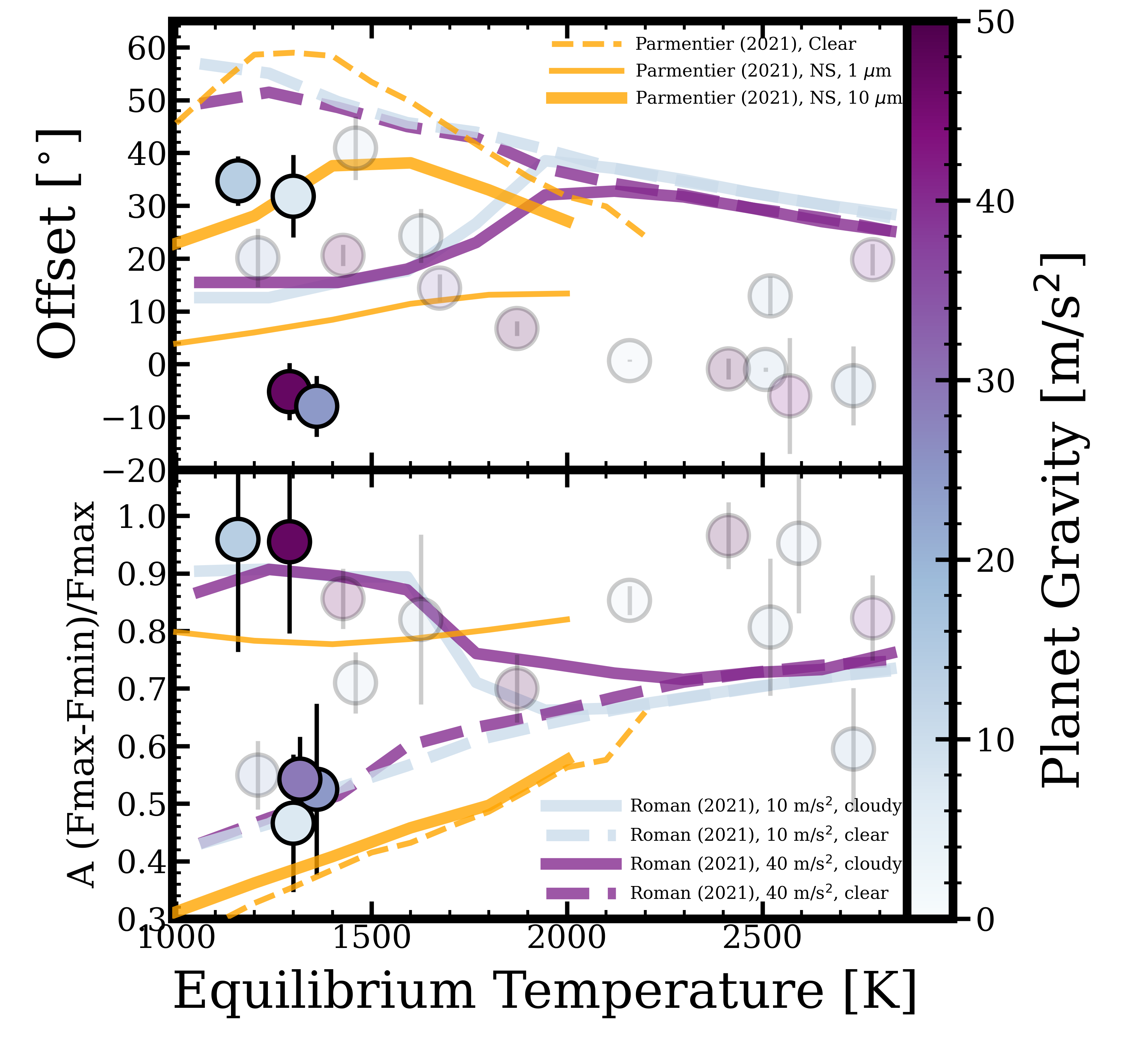}
    \caption{Offset and A$_{obs}$ as a function of Equilibrium temperature for the 4.5 $\micron$ data. It is not always clear if max and min or day and night (phase 1.0/0.5) values are reported in the literature, but we include all literature values here for completeness. Models are from \cite{Roman2021} and \cite{Parmentier2021}. Our new $\sim$1300 K sample matches well with predicted trends from 3D models. We identify a tenative trend of offset and gravity, which we explore more in Figure \ref{fig:offset_v_grav}.}
    \label{fig:romanmodels}
\end{figure}

We also explore previously predicted trends of offset and  relative amplitude (here (F$_{\rm{max}}$ - F$_{\rm{min}}$) / F$_{\rm{max}}$) as a function of equilibrium temperature. Figure \ref{fig:romanmodels} shows the 4.5 $\micron$ data compared to 3D model predictions from \cite{Roman2021} for both cloudy and clear cases at two different planetary gravities. While these models do use radiatively active clouds, which are important at cooler temperatures, we note that these do not consider magnetic drag \citep[see e.g.][]{Rauscher2013,Rogers2014, RogersShowman2014, Beltz2021} which is likely to be important at high temperatures, nor H$_2$ dissociation or non-grey impacts, all of which impact the hotspot offset and circulation \added{\citep[for the impact of the choice of radiative scheme on predicted phase curves see e.g.][]{LeeE2021}. Non-equilibrium chemistry can also impact the emergent phase curve and is not considered in the model predictions we consider here \citep[e.g.][]{Steinrueck2021}.} We also compare to 3D model predictions from \cite{Parmentier2021} for clear and two different nightside (NS) cloud particle sizes. We find good agreement between the \cite{Roman2021} model predictions and our observed amplitudes for our $\sim$ 1300 K sample, but a systematic offset towards smaller hotspot phase offsets for our sample compared to the literature values. However, the \cite{Parmentier2021} model offsets are a better match to our data. The large spread in model predictions highlights the numerous parameters than can impact the shape of hot Jupiter phase curves, as well as the different assumptions involved in these models. 

\begin{figure}
    \centering
    \includegraphics[width=0.5\textwidth]{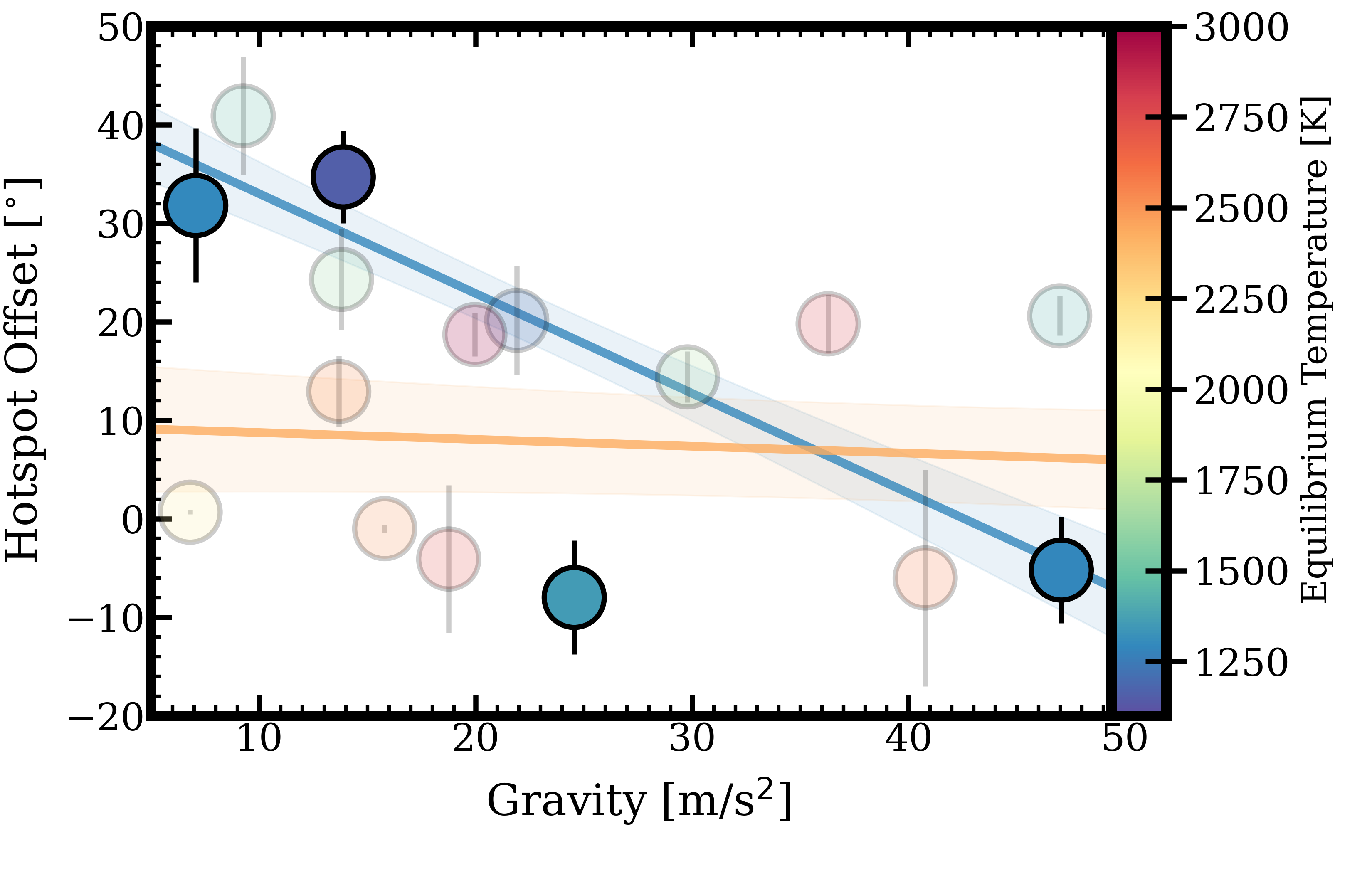}
    \caption{Hotspot offset vs. gravity, colored by equilibrium temperature, for the 4.5 $\micron$ data. Trend lines for our $\sim$1300K sample (blue) and the remaining population (red) are shown. This suggests tentative evidence for a dependence off hotspot offset with gravity for cooler planets, which flattens off at higher equilibrium temperatures.}
    \label{fig:offset_v_grav}
\end{figure}

In addition, there appears to be a slight relationship between gravity and offset seen in our $\sim$1300 K sample (the top panel). We explore this in Figure \ref{fig:offset_v_grav} where we consider trends within our sample and the remaining population (cool planets and warmer planets). There is evidence for a trend within our $\sim$1300 K sample of larger eastward offsets at low gravity, with slight westward offsets at high gravity, with this relationship flattening off for higher temperature planets. This may be expected due to, for example, magnetic drag in ultra hot planets resulting in no offset regardless of gravity. Further modeling including this effect may be useful to explore this trend more, particularly due to the apparent departure between the 10 m/s$^{2}$ and 40 m/s$^{2}$ case in the \cite{Roman2021} models in Figure \ref{fig:romanmodels}.

Finally, our T$_{eq}\sim$ 1300 K sample broadly agrees with previous observations that nightside temperatures of hot Jupiters are all approximately 1000 K \citep{Beatty2019,Keating2019}.

\section{Conclusions} \label{conclusions}
We have presented the analysis of seven new \spitzer\ phase curves, and a reanalysis of one previously published phase curve. The five planets in our sample all have equilibrium temperatures of $\sim$ 1300 K. The analysis of these eight phase curves was performed as uniformly as the data allows with a goal of enabling comparative studies. 

We identify a statistically significant trend of increasing phase curve hotspot offset with orbital period in both our 1300 K sample and the full \spitzer\ sample. Initial comparisons to models of apparent heat redistribution suggest there may be two populations of observed hot Jupiters with weak and strong redistribution, but a more uniform analysis is necessary to definitely make this assessment. We also identify tentative evidence of a hotspot offset dependence on gravity for cool planets. Our newly reduced sample agrees with expectations that planets of this equilibrium temperature should have fairly consistent nightside temperatures near $\sim$ 1000 K. Future work will reanalyze previous Spitzer phase curves with updated and uniform approaches to better understand the scatter in observed trends. 

We also note that it may be informative to compare observational results to model-predicted trends in non-dimensional units that incorporate a combination of parameters (e.g. the ratio of wave propagation and radiative timescales, the non-dimensional Rossby deformation radius), but leave this to future work. With the continued re-analysis of \spitzer\ phase curves we will be able understand how multiple parameters (e.g. rotation rate/orbital period, gravity, equilibrium temperature, etc.) work together to shape the circulation of hot Jupiters, rather than single parameter relationships.

\software{\\ Astropy \citep{astropy,astropy2},
\\ batman \citep{Kreidberg2015},
\\ ExoCTK \citep{exoctk},
\\ IPython \citep{ipython},
\\ Matplotlib \citep{matplotlib},
\\ NumPy \citep{numpy, numpynew},
\\ SciPy \citep{scipy},
}

\acknowledgments
This research has made use of the NASA Exoplanet Archive, which is operated by the California Institute of Technology, under contract with the National Aeronautics and Space Administration under the Exoplanet Exploration Program. This paper makes use of data from the first public release of the WASP data as provided by the WASP consortium and services at the NASA Exoplanet Archive. 

E.M.M.\ acknowledges support from JHU APL's Independent Research And Development program and NASA XRP grant 80NSSC22K0313. K.B.S.\ and J.L.B.\ acknowledge support for this work from NASA through awards issued by JPL/Caltech (Spitzer programs 13038 and 14059). T.J.B. acknowledges support from the McGill Space Institute Graduate Fellowship, the Natural Sciences and Engineering Research Council of Canada’s Postgraduate Scholarships-Doctoral Fellowship, and from the Fonds de recherche du Québec – Nature et technologies through the Centre de recherche en astrophysique du Québec. L.D. acknowledges support in part through the Technologies for Exo-Planetary Science (TEPS) PhD Fellowship, and the Natural Sciences and Engineering Research Council of Canada (NSERC)'s Postgraduate Scholarships-Doctoral Fellowship. J.M.D acknowledges support from the Amsterdam Academic Alliance (AAA) Program, and the European Research Council (ERC) European Union’s Horizon 2020 research and innovation programme (grant agreement no. 679633; Exo-Atmos). This work is part of the research programme VIDI New Frontiers in Exoplanetary Climatology with project number 614.001.601, which is (partly) financed by the Dutch Research Council (NWO). M.M. was supported by NASA through the NASA Hubble Fellowship grant HST-HF2-51485.001-A awarded by the Space Telescope Science Institute, which is operated by the Association of Universities for Research in Astronomy, Inc. under NASA contract NAS 5-26555. Part of the research was carried out at the Jet Propulsion Laboratory, California Institute of Technology, under contract with the National Aeronautics and Space Administration

\clearpage
\appendix
\restartappendixnumbering
\section{Fit Parameters} \label{appendix}
Table \ref{table:fit_params} lists the best fit parameters and their uncertainties (if relevant) for all phase curves. 

\begin{deluxetable}{| r l | c | c | c | c | c | c | c | c |}[h]
    \tablecolumns{10}
    \tabletypesize{\tiny}
    \tablecaption{Fit Parameters}
    \label{table:fit_params}
    \tablehead{
        \colhead{Parameter}&
        \colhead{(Units)} &
        \colhead{qa001bo21} &
        \colhead{qa002bo11} &
        \colhead{qa002bo21} &
        \colhead{wa052bo11} &
        \colhead{wa052bo12} &
        \colhead{wa052bo21} &
        \colhead{wa034bo21} &
        \colhead{wa140bo21} 
    }
    \startdata
    \hline
Transit Midpoint         & \vtop{\hbox{\strut BJD$_{TDB}$}\hbox{\strut -2458000}} & \vtop{\hbox{\strut 242.0175}\hbox{\strut $\pm$ 0.000182}} &  \vtop{\hbox{\strut -95.9549}\hbox{\strut $\pm$ 9.4e-05}} &  \vtop{\hbox{\strut -103.9772}\hbox{\strut $\pm$ 0.000183}} &  \vtop{\hbox{\strut -320.0543}\hbox{\strut $\pm$ 0.000114}} &  \vtop{\hbox{\strut -320.0543}\hbox{\strut $\pm$ 0.000114}} &  \vtop{\hbox{\strut 199.6311}\hbox{\strut $\pm$ 0.00014}} &  \vtop{\hbox{\strut 762.307}\hbox{\strut $\pm$ 0.0002}} &  \vtop{\hbox{\strut 486.4851}\hbox{\strut $\pm$ 0.000123}} \\ \hline 
Rp/Rs                    &            - & \vtop{\hbox{\strut 0.1451}\hbox{\strut $\pm$ 0.000823}} &  \vtop{\hbox{\strut 0.1637}\hbox{\strut $\pm$ 0.000519}} &  \vtop{\hbox{\strut 0.1619}\hbox{\strut $\pm$ 0.001007}} &  \vtop{\hbox{\strut 0.1646}\hbox{\strut $\pm$ 0.000606}} &  \vtop{\hbox{\strut 0.1646}\hbox{\strut $\pm$ 0.000606}} &  \vtop{\hbox{\strut 0.163}\hbox{\strut $\pm$ 0.000721}} &  \vtop{\hbox{\strut 0.1201}\hbox{\strut $\pm$ 0.001122}} &  \vtop{\hbox{\strut 0.174}\hbox{\strut $\pm$ 0.000689}} \\ \hline 
Period                   &         Days & 1.42& 1.3371& 1.3371& \vtop{\hbox{\strut 1.7498}\hbox{\strut $\pm$ 4e-06}} &  \vtop{\hbox{\strut 1.7498}\hbox{\strut $\pm$ 4e-06}} &  1.7498& 4.3177& 2.2369\\ \hline
a/Rs                     &            - & 6.247& \vtop{\hbox{\strut 5.96}\hbox{\strut $\pm$ 0.013589}} &  \vtop{\hbox{\strut 5.9641}\hbox{\strut $\pm$ 0.021705}} &  \vtop{\hbox{\strut 7.265}\hbox{\strut $\pm$ 0.010448}} &  \vtop{\hbox{\strut 7.265}\hbox{\strut $\pm$ 0.010448}} &  7.1989& \vtop{\hbox{\strut 10.6878}\hbox{\strut $\pm$ 0.028778}} &  7.98\\ \hline
cos(i)                   &            - & 0.1031& 0.0677& 0.0677& 0.0811& 0.0811& 0.0811& 0.0837& 0.1156\\ \hline
e                        &            - & 0.0& 0.0& 0.0& 0.0& 0.0& 0.0& 0.0& 0.047\\ \hline
$\Omega$                 &   $^{\circ}$ & 0.0& 0.0& 0.0& 0.0& 0.0& 0.0& 0.0& -4.0\\ \hline
u$_1$                    & (limb dark.) & 0.101& 0.1116& 0.1& 0.108& 0.108& 0.108& 0.09& 0.1\\ \hline
u$_2$                    & (limb dark.) & 0.113& 0.1166& 0.125& 0.147& 0.147& 0.147& 0.099& 0.125\\ \hline
Eclipse Midpoint 1       & \vtop{\hbox{\strut BJD$_{TDB}$}\hbox{\strut -2458000}} & \vtop{\hbox{\strut 241.3086}\hbox{\strut $\pm$ 0.001053}} &  \vtop{\hbox{\strut -96.6226}\hbox{\strut $\pm$ 0.000844}} &  \vtop{\hbox{\strut -104.6464}\hbox{\strut $\pm$ 0.0015}} &  \vtop{\hbox{\strut -320.9297}\hbox{\strut $\pm$ 0.001204}} &  \vtop{\hbox{\strut 48.2759}\hbox{\strut $\pm$ 0.001246}} &  \vtop{\hbox{\strut 198.7568}\hbox{\strut $\pm$ 0.000892}} &  \vtop{\hbox{\strut 760.146}\hbox{\strut $\pm$ 0.00242}} & \vtop{\hbox{\strut 485.4285}\hbox{\strut $\pm$ 0.0028}} \\ \hline 
T$_{14}$                 &         Days & 0.0672& 0.0754& 0.0754& 0.0754& 0.0754& 0.0754& - & 0.0631\\ \hline
Eclipse Depth            &          ppm & \vtop{\hbox{\strut 2900.0}\hbox{\strut $\pm$ 161.0}} &  \vtop{\hbox{\strut 2200.0}\hbox{\strut $\pm$ 118.0}} &  \vtop{\hbox{\strut 3000.0}\hbox{\strut $\pm$ 185.0}} &  \vtop{\hbox{\strut 2200.0}\hbox{\strut $\pm$ 77.0}} &  \vtop{\hbox{\strut 2200.0}\hbox{\strut $\pm$ 77.0}} &  \vtop{\hbox{\strut 3400.0}\hbox{\strut $\pm$ 162.0}} &  - & \vtop{\hbox{\strut 2000.0}\hbox{\strut $\pm$ 111.0}} \\ \hline 
T$_{12}$                 &         Days & 0.0142& 0.0107& 0.0107& 0.0076& 0.0076& 0.0076& - & \vtop{\hbox{\strut 0.0223}\hbox{\strut $\pm$ 0.004917}} \\ \hline 
T$_{34}$                 &         Days & 0.0142& 0.0107& 0.0107& 0.0076& 0.0076& 0.0076& - & \vtop{\hbox{\strut 0.0135}\hbox{\strut $\pm$ 0.006333}} \\ \hline 
Eclipse Midpoint 2       & \vtop{\hbox{\strut BJD$_{TDB}$}\hbox{\strut -2458000}} & \vtop{\hbox{\strut 242.7272}\hbox{\strut $\pm$ 0.001236}} &  \vtop{\hbox{\strut -95.2854}\hbox{\strut $\pm$ 0.001152}} &  \vtop{\hbox{\strut -103.3064}\hbox{\strut $\pm$ 0.001081}} &  \vtop{\hbox{\strut -319.1775}\hbox{\strut $\pm$ 0.000764}} &  \vtop{\hbox{\strut 50.0266}\hbox{\strut $\pm$ 0.001014}} &  \vtop{\hbox{\strut 200.5069}\hbox{\strut $\pm$ 0.000744}} &  $^{\dagger}$ & \vtop{\hbox{\strut 487.664}\hbox{\strut $\pm$ 0.003004}} \\ \hline 
Fp/Fs$^{\dagger}$                    &          ppm & - & - & - & - & - & - & \vtop{\hbox{\strut 800.0}\hbox{\strut $\pm$ 97.0}} &  - \\ \hline
Full Per. Cos Amp.       &          ppm & \vtop{\hbox{\strut 800.0}\hbox{\strut $\pm$ 144.0}} &  \vtop{\hbox{\strut 1000.0}\hbox{\strut $\pm$ 141.0}} &  \vtop{\hbox{\strut 1400.0}\hbox{\strut $\pm$ 184.0}} &  \vtop{\hbox{\strut 600.0}\hbox{\strut $\pm$ 65.0}} &  \vtop{\hbox{\strut 600.0}\hbox{\strut $\pm$ 65.0}} &  \vtop{\hbox{\strut 800.0}\hbox{\strut $\pm$ 180.0}} &  \vtop{\hbox{\strut 400.0}\hbox{\strut $\pm$ 41.0}} &  \vtop{\hbox{\strut 600.0}\hbox{\strut $\pm$ 24.0}} \\ \hline 
Full Per. Cos Offset     &         Days & \vtop{\hbox{\strut 0.7601}\hbox{\strut $\pm$ 0.032978}} &  \vtop{\hbox{\strut 0.7122}\hbox{\strut $\pm$ 0.021513}} &  \vtop{\hbox{\strut 0.7101}\hbox{\strut $\pm$ 0.019904}} &  \vtop{\hbox{\strut 0.8301}\hbox{\strut $\pm$ 0.023567}} &  \vtop{\hbox{\strut 0.8301}\hbox{\strut $\pm$ 0.023567}} &  \vtop{\hbox{\strut 0.852}\hbox{\strut $\pm$ 0.063766}} &  \vtop{\hbox{\strut 2.0454}\hbox{\strut $\pm$ 0.05615}} &  \vtop{\hbox{\strut 1.7087}\hbox{\strut $\pm$ 0.021923}} \\ \hline 
Half Per. Cos Amp.       &          ppm & - & - & - & - & - & - & - & \vtop{\hbox{\strut 200.0}\hbox{\strut $\pm$ 32.0}} \\ \hline 
Half Per. Cos Offset     &         Days & - & - & - & - & - & - & - & \vtop{\hbox{\strut 0.0607}\hbox{\strut $\pm$ 0.027693}} \\ \hline 
Quadratic Ramp Term      &            - & \vtop{\hbox{\strut -0.0019}\hbox{\strut $\pm$ 0.000485}} &  \vtop{\hbox{\strut -0.0026}\hbox{\strut $\pm$ 0.0006}} &  \vtop{\hbox{\strut 0.0005}\hbox{\strut $\pm$ 0.000755}} &  - & - & - & - & - \\ \hline
Linear Ramp Term         &            - & \vtop{\hbox{\strut 0.0024}\hbox{\strut $\pm$ 0.000241}} &  \vtop{\hbox{\strut 0.0023}\hbox{\strut $\pm$ 0.000229}} &  \vtop{\hbox{\strut 0.0012}\hbox{\strut $\pm$ 0.000226}} &  \vtop{\hbox{\strut 0.0032}\hbox{\strut $\pm$ 7.9e-05}} &  \vtop{\hbox{\strut 0.0187}\hbox{\strut $\pm$ 1.8e-05}} &  - & - & - \\ \hline
Ramp Offset              &         Days & 0.0& 0.0& 0.0& 0.0& 4.0& - & - & - \\ \hline
PRF, Linear X            &            - & - & \vtop{\hbox{\strut 0.0683}\hbox{\strut $\pm$ 0.003633}} &  \vtop{\hbox{\strut -0.0009}\hbox{\strut $\pm$ 0.005045}} &  \vtop{\hbox{\strut -0.164}\hbox{\strut $\pm$ 0.005081}} &  \vtop{\hbox{\strut -0.007}\hbox{\strut $\pm$ 0.004743}} &  \vtop{\hbox{\strut 0.0121}\hbox{\strut $\pm$ 0.005777}} &  - & - \\ \hline
PRF, Quad. X             &            - & - & \vtop{\hbox{\strut 0.0572}\hbox{\strut $\pm$ 0.004632}} &  \vtop{\hbox{\strut 0.2461}\hbox{\strut $\pm$ 0.057581}} &  \vtop{\hbox{\strut 0.1696}\hbox{\strut $\pm$ 0.040813}} &  \vtop{\hbox{\strut -0.4528}\hbox{\strut $\pm$ 0.07063}} &  \vtop{\hbox{\strut 0.0598}\hbox{\strut $\pm$ 0.035102}} &  - & - \\ \hline
PRF, Linear Y            &            - & - & \vtop{\hbox{\strut -0.107}\hbox{\strut $\pm$ 0.004816}} &  \vtop{\hbox{\strut -0.2576}\hbox{\strut $\pm$ 0.016666}} &  \vtop{\hbox{\strut -0.1258}\hbox{\strut $\pm$ 0.007046}} &  \vtop{\hbox{\strut -0.0818}\hbox{\strut $\pm$ 0.006162}} &  \vtop{\hbox{\strut 0.0336}\hbox{\strut $\pm$ 0.005722}} &  - & - \\ \hline
PRF, Quad. Y             &            - & - & \vtop{\hbox{\strut -0.1499}\hbox{\strut $\pm$ 0.051506}} &  \vtop{\hbox{\strut -2.5815}\hbox{\strut $\pm$ 0.134902}} &  \vtop{\hbox{\strut -1.34}\hbox{\strut $\pm$ 0.090374}} &  \vtop{\hbox{\strut 0.3641}\hbox{\strut $\pm$ 0.029821}} &  \vtop{\hbox{\strut -0.2011}\hbox{\strut $\pm$ 0.033294}} &  - & - \\ \hline
PRF Offset               &     PRF FWHM & - & 0.6& 0.6& 0.6& 0.6& 0.6& - & - \\ \hline
Constant                 &   Flux units & \vtop{\hbox{\strut 10716.2499}\hbox{\strut $\pm$ 1.117601}} &  \vtop{\hbox{\strut 13813.3694}\hbox{\strut $\pm$ 6.325889}} &  \vtop{\hbox{\strut 8547.7662}\hbox{\strut $\pm$ 4.977395}} &  \vtop{\hbox{\strut 24419.5982}\hbox{\strut $\pm$ 1.055123}} &  \vtop{\hbox{\strut 2977.2732}\hbox{\strut $\pm$ 2.310508}} &  \vtop{\hbox{\strut 14058.807}\hbox{\strut $\pm$ 7.043908}} &  \vtop{\hbox{\strut 49752.8256}\hbox{\strut $\pm$ 4.53561}} &  \vtop{\hbox{\strut 34490.6782}\hbox{\strut $\pm$ 0.564251}} \\ \hline 
    \enddata
    \tablecomments{Parameters in fits. Those without errors are held constant. All parameters have uniform priors that span the entire width of physically possible values.
    \\ $^{\dagger}$ Due to the grazing nature of the WASPS-34b occultations we use a different functional form for the eclipses, as described in the text. This is a repeating function, so there is no second midpoint. The Fp/Fs term is also unique to this event and is comparable to the eclipse depth. The values for WASP-34b in this table are not corrected to account for the grazing event.}
\end{deluxetable}

\bibliography{REFS}{}
\bibliographystyle{aasjournal}

\end{document}